\documentclass{aa}  
\usepackage{natbib}
\usepackage{graphicx}
\usepackage{txfonts}
\usepackage{hyperref}
\usepackage{color}
\newcommand{\uv}{$\{u, \mathrm{v}\}$}

\newcommand{\modif}[1]{#1}

\newcommand{\denv}{\ensuremath{d_\mathrm{env}}\xspace}
\begin{document} 

  \title{The perturbed sublimation rim of the dust disk around the post-AGB binary IRAS08544-4431\thanks{Based on observations performed with PIONIER mounted on the ESO \textit{Very Large Telescope interferometer} (programme: 094.D-0865).}}
 
  \author{J. Kluska\inst{1}
     \and
           M. Hillen\inst{1}
            \and
           H. Van Winckel\inst{1}
           \and
           R. Manick\inst{1}
           \and
           M. Min\inst{2,3}
           \and
           S. Regibo\inst{1}
           \and
           P. Royer\inst{1}
          }

  \institute{Instituut voor Sterrenkunde (IvS), KU Leuven, Celestijnenlaan 200D, 3001 Leuven, Belgium\\
              \email{jacques.kluska@kuleuven.be}
              \and
    SRON Netherlands Institute for Space Research, Sorbonnelaan 2, 3584 CA, Utrecht, The Netherlands
    \and 
    Astronomical institute Anton Pannekoek, University of Amsterdam, Science Park 904, 1098 XH, Amsterdam, The Netherlands
              }

  \date{Received February 28, 2018; accepted}

  \abstract
   {Post-Asymptotic Giant Branch (AGB) binaries are surrounded by stable dusty and gaseous disks similar to the ones around young stellar objects. 
   Whereas significant effort is spent on modeling observations of disks around young stellar objects, the disks around post-AGB binaries receive significantly less attention, even though they pose significant constraints on theories of disk physics and binary evolution.}
   {We want to examine the structure of and phenomena at play in circumbinary disks around post-AGB stars.
   We continue the analysis of our near-infrared interferometric image of the inner rim of the circumbinary disk around IRAS08544-4431. 
   We want to understand the physics governing this inner disk rim.}
   {We use a radiative transfer model of a dusty disk to reproduce simultaneously the photometry as well as the near-infrared interferometric dataset on IRAS08544-4431.
   The model assumes hydrostatic equilibrium and takes dust settling self-consistently into account. }
   {The best-fit radiative transfer model shows excellent agreement with the spectral energy distribution up to mm wavelengths as well as with the PIONIER visibility data. 
   It  requires a rounded inner rim structure, starting at a radius of 8.25\,au.
   However, the model does not fully reproduce the detected over-resolved flux nor the azimuthal flux distribution of the inner rim. 
   While the asymmetric inner disk rim structure is likely to be the consequence of disk-binary interactions, the origin of the additional over-resolved flux remains unclear.}
   {As in young stellar objects, the disk inner rim of IRAS08544-4431 is ruled by dust sublimation physics.
   Additional observations are needed to understand the origin of the extended flux and the azimuthal perturbation at the inner rim of the disk.}


\maketitle

%________________________________________________________________

\section{Introduction}

It is by now well established that some post-asymptotic giant branch (post-AGB) binaries are surrounded by stable circumbinary disks \citep[e.g.][]{vanwinckel03,vanwinckel17}. Observational signatures are the presence of a flux excess at near-infrared (near-IR) wavelengths, indicating that this circumstellar dust must be close to the
central star near sublimation temperatures, while the central stars are not in a dust-losing phase \citep[e.g.][]{deruyter06}. 
A typical picture of a post-AGB binary involves that the unevolved star hosts an accretion disk and an outflow.
The inner binary system is surrounded by a circum-binary disk that is possibly accreted onto the inner binary.

Observational indicators for longevity of the disk are the evidence of strong dust grain processing in the form of a high degree of crystallinity 
\citep[e.g.][]{gielen08,gielen11} and the presence of large grains \citep[e.g.][]{deruyter05, gielen11,2015aahillen} in many systems. 
The strongest observational evidence comes from the objects in which the Keplerian velocity field is spatially resolved in CO using the ALMA and the Plateau De Bure interferometers \citep{bujarrabal13a,bujarrabal15,bujarrabal17a}. 
The single-dish survey of \citet{bujarrabal13b} confirms that rotation must be widespread among these sources. 

As the infrared emission is compact, interferometric techniques are needed to resolve the inner dusty region. Applications of high-spatial-resolution techniques \citep{deroo06,deroo07,2013aahillen,2014aahillen,2015aahillen,hillen17} unveiled both the compact inner region as well as the strong similarity between these disks and protoplanetary disks in hydrostatic equilibrium.
The infrared luminosity of these objects is significant, pointing to a large scaleheight of the disk. 
The near-IR excess is originating specifically from the inner rim of the disk.

Based on the spectral energy distribution (SED), new
such objects can be efficiently identified.  
In recent searches for post-AGB stars in the Large and Small Magellanic Clouds \citep{vanaarle11,kamath14a, kamath15}, disk sources represent about half of the population of optically bright post-AGB stars. 
Disks also appear at lower luminosities, indicating that the central evolved star is a post-Red Giant Branch (post-RGB) star, rather than a post-AGB star \citep{kamath16}.
In the Galaxy a sample of about 85 of these disk objects have been identified \citep{deruyter06,gezer15}.

The orbits determined so far \citep[e.g.][]{vanwinckel09,manick17,oomen18} are too small to accommodate an AGB star. The evolved binary escaped the phase of strong interaction when the primary was at giant dimensions on the AGB on a surprisingly wide and often eccentric orbit. 
While the post-AGB stars are now within their Roche-lobe, there is strong observational evidence for continuous interaction between the disk and the binary. 
Indeed, the post-AGB stars with disk-SEDs display often a chemical anomaly in which the refractory elements are depleted, while the volatile elements have higher abundances. The photospheric abundances scale with the condensation temperature of the chemical element \citep[e.g.][and references therein]{gezer15}. 
While this depletion is not well understood yet, the general picture as described by \cite{Waters1992} is generally acknowledged: circumstellar gas devoid of refractory elements (as these remain part of the dust in the disk) is re-accreted by the post-AGB star. 
The presence of a stable disk is a needed but not a sufficient condition for dust-gas separation and re-accretion of cleaned gas to occur.

The binaries should be seen as still interacting. 
Using orbital phase resolved spectroscopy, several systems are documented with observational evidence for a fast outflow, originating around the companion \citep[e.g.][]{thomas13,gorlova15,bollen17}. 
The accretion disk launches a fast outflow, which is seen in absorption at superior conjunction (i.e. when the secondary is seen in front of the primary). 
The physical model is that continuum photons of the primary are scattered outside the line-of-sight when passing through the jet. 
The measured deprojected outflow velocities are indicative of the escape velocity of a main sequence companion and not a White Dwarf (WD). 
Whether the circum-companion accretion disks are fed by the circumbinary dusty disk, or the evolved primary is not yet known.

The stable circumbinary dusty disks are therefore thought to play a lead role in the final evolution of a large population of binary stars. 
However, their structure, dispersal and evolution remain elusive. 
We therefore started a large project to study the physical processes which govern the very inner region of these systems on the basis of dedicated multi-wavelength and high spatial-resolution observations.

In this contribution we report on a very specific system IRAS08544-4431. 
In our earlier paper \citep[][hereafter Paper~I]{Hillen2016}, we presented the results of our successful interferometric imaging experiment using the PIONIER instrument \citep{2011aalebouquin} on the Very Large Telescope Interferometer (VLTI) of ESO. 
The inner rim of the disk is well resolved and the different components contributing to the flux in the H-band are: the central star, the accretion disk around the companion, the inner rim of the circumbinary disk and an overresolved scattering component. 
While the near-IR is sensitive to the very inner region of the disk which determines the energetics of the system, the outer region of the disk of IRAS08544-4431 was resolved in CO (3-2) using the ALMA array \citep{Bujarrabal2018}. A complex resolved profile emerged and different regions could be discriminated: the inner gaseous disk, likely with a molecular free inner region has a radius of around 600\,au and was found to be in Keplerian rotation, while the extended disk reaching a radius of about 1300\,au was found to be at sub-Keplerian velocities. On top of that an even larger CO slow outflow was resolved.

In this contribution we present the physical model of the circumbinary disk based on 2D radiative transfer modeling and we focus on the very inner parts. 
We also perform an image reconstruction of the best-fit model in order to compare it to the image from the actual interferometric dataset of IRAS~08544-4431.
The paper is organised as follows: in Sect.\,\ref{sec:obs} we describe the observations that we analyse in Sect.\,\ref{sec:ana}.
Then, we discuss our findings in Sect.\,\ref{sec:dis} and conclude in Sect.\,\ref{sec:ccl}.

%__________________________________________________________________

\section{Observations}
\label{sec:obs}

\subsection{Photometry}

To constrain the overall energetics of the object, we assembled broad-band photometric data to construct the full spectral energy distribution (SED). 
From published catalogs we collected measurements in several photometric bandpasses (see Appendix\,\ref{app:sed}). 
In addition, we include new SPIRE \citep{spiregriffin} photometry taken with the Herschel satellite \citep{2010aapilbratt}. 
The observations were done simultaneously in three bands (250, 350 and 500\,$\mu$m) on December 22nd 2012 (obs. id. 1342247249; program OT2\_cgielen\_4). 
The fluxes are extracted with the standard 'timeline extraction' method. The uncertainties on the fluxes are dominated by the absolute flux calibration \citep[we assume the upper limit of 15\%;][]{spireswinyard}.  
We also include a sub-mm flux measurement, acquired with the LABOCA \citep{Siringo2009} instrument that is mounted on the APEX telescope \citep{Gusten2006}. 
This 295-bolometer total power camera was used on October 23rd 2008 to observe the continuum emission at 870\,$\mu$m (4 scans with 35\,s integration time in OTF mode). 
Standard procedures were applied for the data reduction and calibration.

\subsection{Extended CORALIE dataset}

\begin{figure}[th]
    \centering
    \includegraphics[width=8cm]{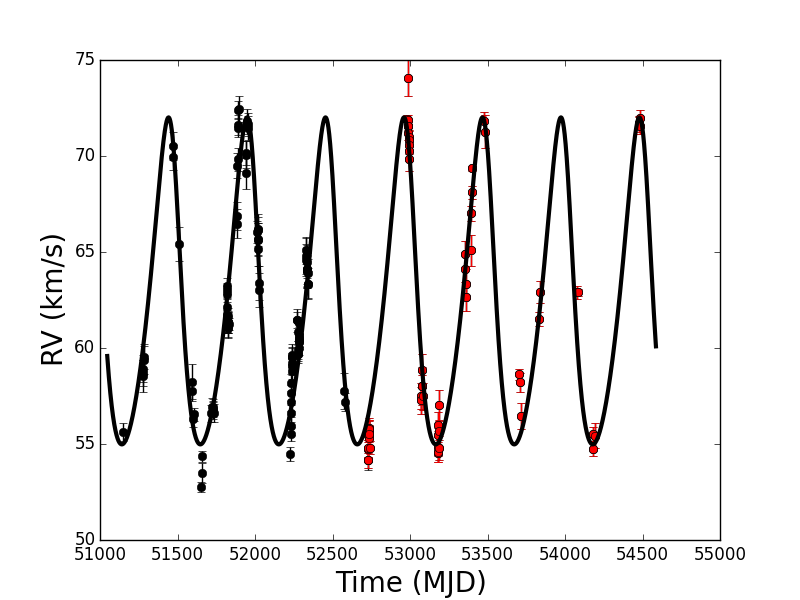}
    \caption{Radial velocity time series with the best-fit orbital model (the full black line). Black circles are measurements that were previously published. Red circles represent new data.}
    \label{fig:CORALIE}
\end{figure}

We extended the spectroscopic time series of IRAS08544-4431 \citep{2003aamaas} with more data from the Swiss 1.2m Telescope at La Silla, on which the CORALIE spectrograph \citep{Queloz1999} is mounted (see Fig.\,\ref{fig:CORALIE}). 
The extended time base of the radial velocities allows a more accurate determination of the spectroscopic orbital elements. 
IRAS08544-4431 displays complex light variations, with several low-amplitude pulsation modes being excited \citep{2007mnraskiss}. 
The radial velocities are not only affected by orbital motion, but also by these complex photospheric variations.

\subsection{PIONIER}

%                                     Two column figure (place early!)
%______________________________________________ Gamma_1 (lg rho, lg e)
\begin{figure}
    \centering
    \includegraphics[width=5.5cm]{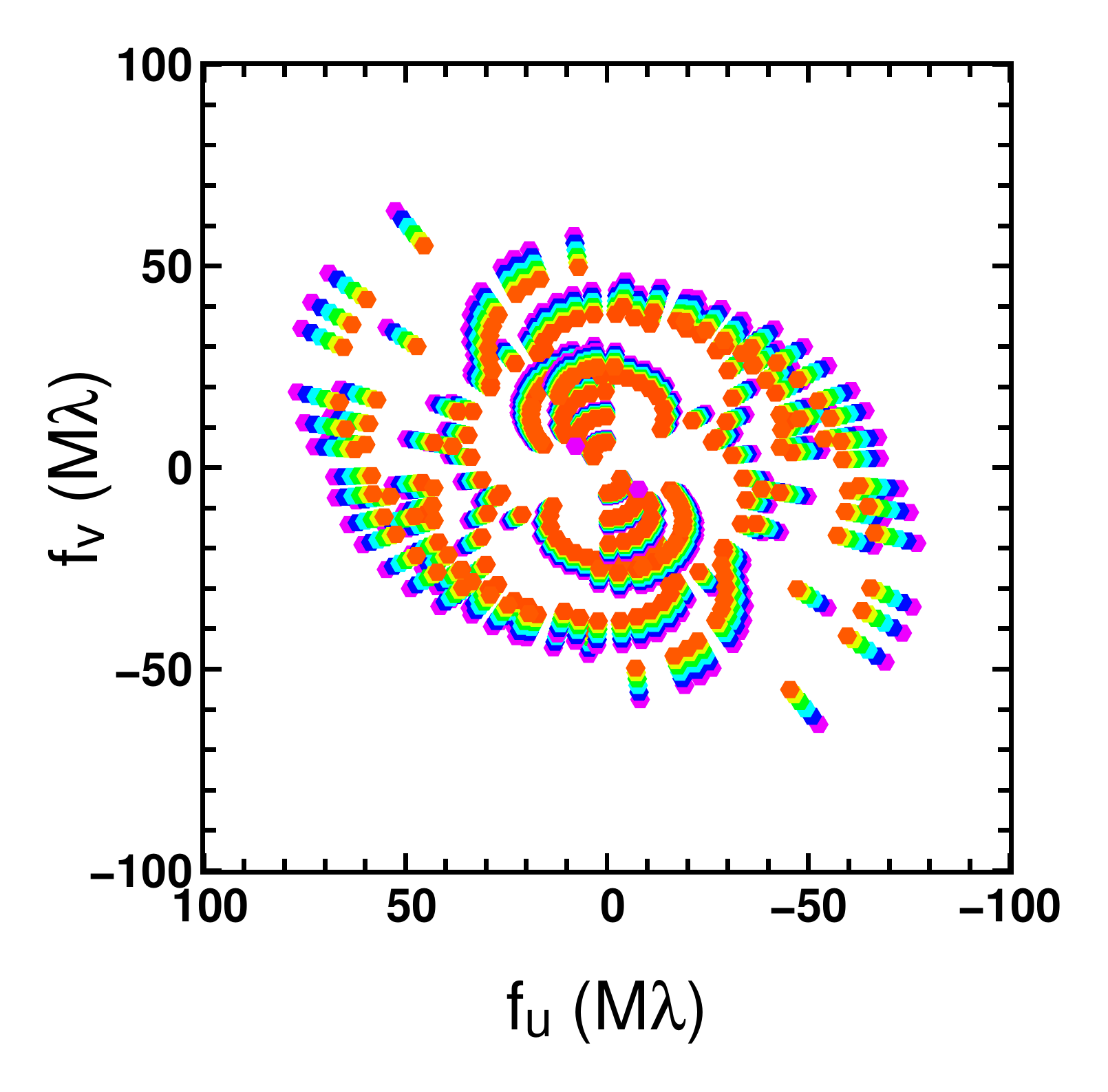}\\
    \includegraphics[width=4.4cm]{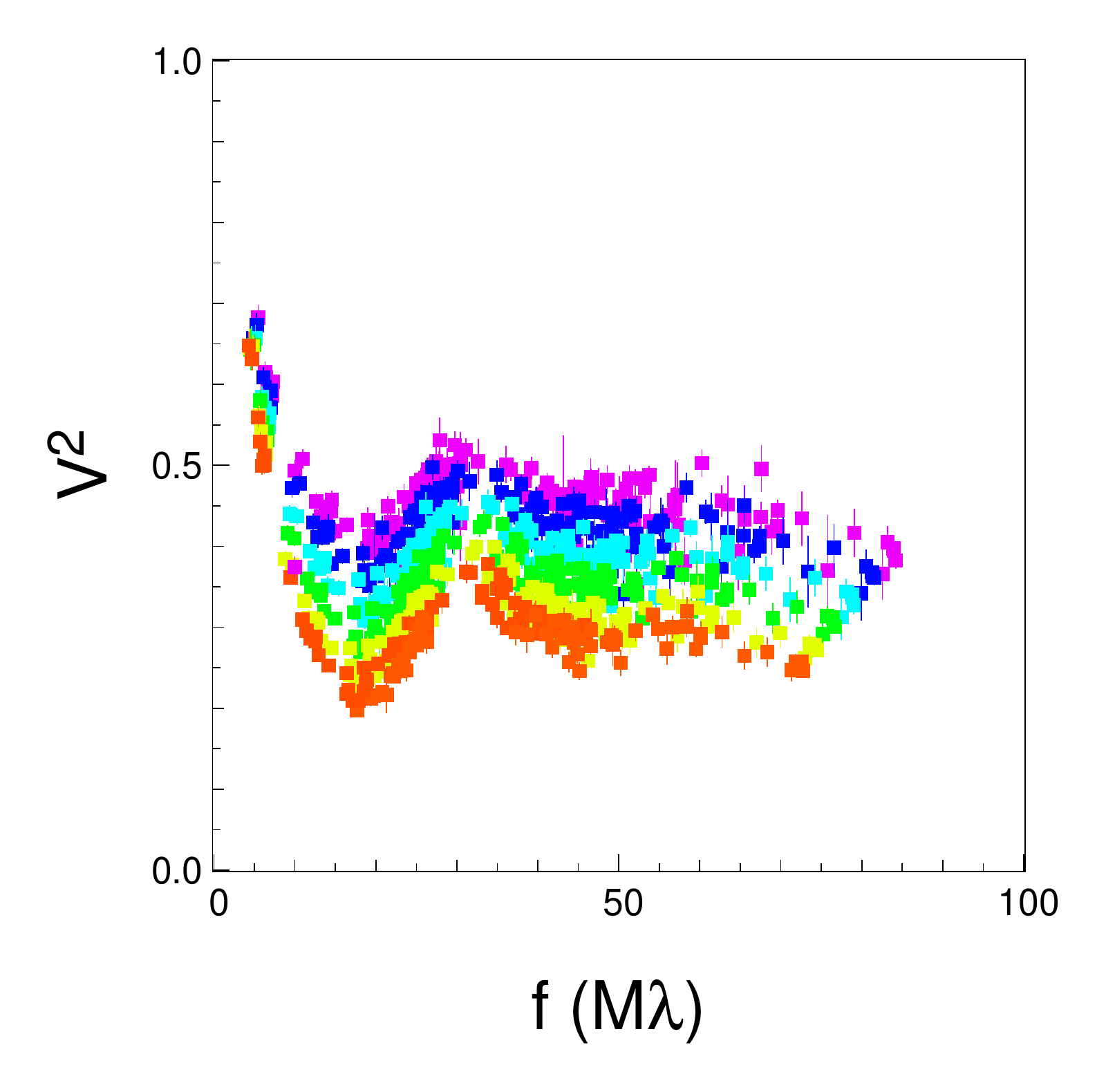}
    \includegraphics[width=4.4cm]{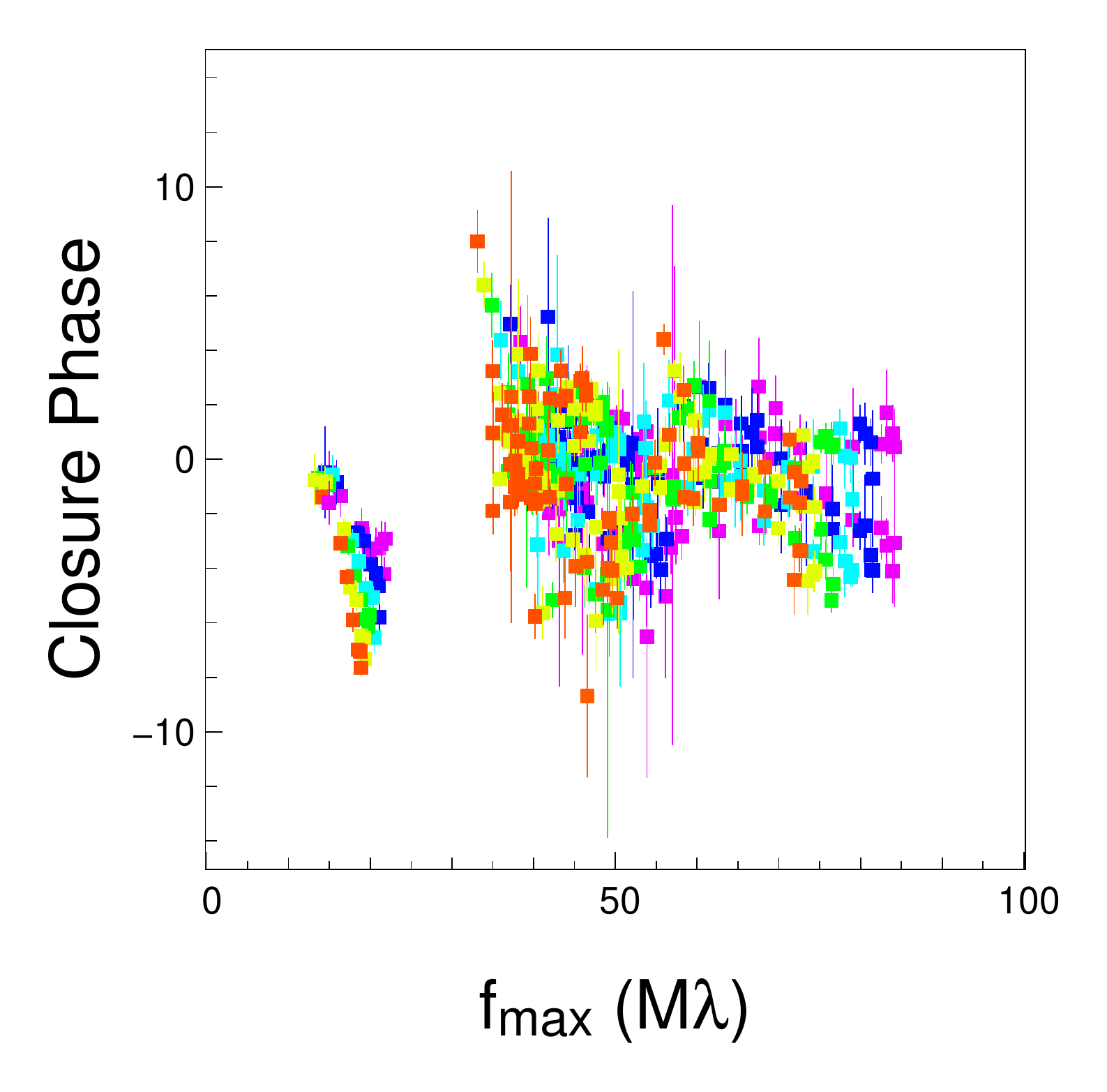}
    \caption{Interferometric dataset of IRAS08544-4431. Top: the \uv-coverage. Left: The V$^2$ versus spatial frequency ($f$). Right: the closure phases versus the maximum baseline spatial frequency ($f_\mathrm{max}$). The color represent the wavelengths (red: 1.85$\mu$m, blue: 1.55$\mu$m).}
    \label{fig:PIONIER}%
\end{figure}

The interferometric dataset was taken with the PIONIER instrument \citep{2011aalebouquin} which is a four-beam combiner mounted on the VLTI.
This instrument operates in the near-infrared $H$-band (1.65$\mu$m).
The dataset was taken on the nights of 2015-01-21, 2015-01-24 and 2015-02-23 (prog. ID: 094.D-0865, PI: Hillen).
Here we remind its main characteristics.
It consists in 828 data points divided in 6 spectral channels accross the $H$-band.
They were recorded on the three auxiliary telescope configurations available at the VLTI resulting in a \uv-coverage with baselines ranging from 7 to 129\,m (see Fig.\,\ref{fig:PIONIER}). 
We refer to Paper~I for a more extensive description of the data.
  
\section{Analysis }
\label{sec:ana}

\begin{table}[!t]
\caption{$H$-band parameters of IRAS08544-4431 inferred from Paper~I.}
\begin{center}
\begin{tabular}{l|c}
Parameter & Value  \\
 \hline
 Flux of the primary \modif{(\%)} & 59.7 $\pm$ 0.6\\
 Flux of the secondary \modif{(\%)}&  3.9 $\pm$ 0.7\\
 Over-resolved flux \modif{(\%)} & 15.5 $\pm$ 0.5 \\
 Binary separation \modif{(mas)}& 0.81 $\pm$ 0.05 \\
 Binary position angle \modif{($^\circ$)}& 56 $\pm$ 3 \\
 Disk inclination \modif{($^\circ$)} & 19 $\pm$ 2 \\
 Disk position angle \modif{($^\circ$)} & 6 $\pm$ 6 \\
 Disk inner radius \modif{(mas)}& 7.56 $\pm$ 0.05
\end{tabular}
%\tablefoot{}
%\tablefoottext{*}{Objects without a regular inclination pattern, see Sect.\,\ref{sec:azimpro}.}}
\end{center}
\label{tab:PIOresults}
\end{table}%

\subsection{Updated spectroscopic orbital elements of the central binary}
  
\begin{table}[!t]
\caption{Updated spectroscopic orbital parameters of IRAS08544-4431.}
\begin{center}
\begin{tabular}{l|c}
Parameter & Value  \\
 \hline
 Orbital period $P$ (days) & 506.0 $\pm$ 1.3\\
 Eccentricity $e$ & 0.22 $\pm$ 0.02 \\
 Argument of periastron $\omega$ (rad) & 0.73 $\pm$ 0.11\\
 Time of periastron passage $T_\mathrm{0}$ (MJD) & 51466 $\pm$ 9\\
 Velocity semi-amplitude $K$ (km/s) & 8.7 $\pm$ 0.2\\
 Systemic velocity $\gamma$ (km/s) & 62.2 $\pm$ 0.1\\
 Primary semi-major axis $a_1 \sin(i)$ (au) & 0.39 $\pm$ 0.01\\
 Mass function $f(m)$ (M$_\odot$) & 0.031 $\pm$ 0.002 
\end{tabular}
%\tablefoot{}
\end{center}
\label{tab:CORALIE}
\end{table}%

To derive the spectroscopic orbital elements, the raw radial velocities were fit with a Keplerian model. 
We iteratively pre-whitened the pulsation signal with a Lomb-Scargle method. 
The dominant pulsation period was subtracted from the original data and the residuals were used to determine the new orbital elements, after which the iteration process was restarted. 
The fractional variance reduction was used as a stop criterion.

We found two dominant pulsation periods in the radial velocity data, 69 d and 77 d, which are in accordance with the 68.9 d and 72.3 d periods found in the light curve \citep{2007mnraskiss}. 
We subtracted the pulsation model with the 69 d and 77 d periods from the original signal and derived the optimized orbital parameters, as listed in Table\,\ref{tab:CORALIE}. 
The uncertainties were computed with a Monte Carlo simulation, assuming Gaussian-distributed noise and making 250 equivalent data sets.
  
We found a period of 506.0 days which is at 2.3-$\sigma$ of the one determined previously in \citet{2003aamaas}.
The eccentricity ($e$) we derived is also higher, 0.22, at 4-$\sigma$ from the one derived previously (0.14).
These discrepancies are likely due to the pulsations that interfer\modif{e} with the binary signal in the radial velocities and which were not taken into account in the earlier analysis.

\subsection{Determination of all the orbital elements of the central binary}

\begin{table}[!t]
\caption{2-$\sigma$ lower limits on fundamental parameters of the binary system, as derived from the combination of interferometric and radial velocity data.}
\begin{center}
\begin{tabular}{l|c}
Parameter & Value  \\
 \hline
 Mass of the primary star $M_1$ (M$_\odot$) &  0.75 \\
Mass of the secondary star $M_2$ (M$_\odot$) & 1.65 \\
Separation $a$ (au) &  1.74 \\
Luminosity of the primary star $L_1$ (L$_\odot$) & 14000 \\
Radius of the primary star $R_1$ (R$_\odot$) & 75 \\
Distance $d$ (kpc) & 1.4
\end{tabular}
%\tablefoot{}
\end{center}
\label{tab:fundamentalparam}
\end{table}%

We now put limits on some physical parameters of the system, i.e. the primary's luminosity and radius, and the distance. 
Often a canonical post-AGB luminosity of 5000\,L$_\odot$ is assumed, from which a distance of $\sim$0.8\,kpc can be derived. 
The Gaia mission found a parallax of 0.86$\pm$0.6\,mas for IRAS08544-4431 \citep{Gaia}. 
This parallax is likely to be affected by the orbital motion as the parallax value is similar to the binary angular separation that we detect in our data (assuming that the off-center point source in our model indicates the position of the secondary star).

Here we use the detected angular separation ($\rho_\mathrm{bin}$) between the two stars (0.8\,mas; Paper~I) to estimate the distance. 
This requires the projected linear separation ($r_\mathrm{proj}$) to be known: $d=r_\mathrm{proj}/\rho_\mathrm{bin}$, with $d$ being the distance to the target.
We compute the projected linear separation at the time of the PIONIER observations with the Thiele-Innes constants \citep{Hilditch2001}:
\begin{equation}
    r_\mathrm{proj}=a \sqrt{[\cos(\omega)X-\sin(\omega)Y]^2+\cos^2(i)[\sin(\omega)X+\cos(\omega)Y]^2}
\end{equation}
 with $a=a_1(M_1+M_2)/M_2$, $X=\cos(E)-e, Y=\sqrt(1-e^2) \sin(E)$ and $E$ the eccentric anomaly in Kepler's equation $E-e\sin(E)=2\pi(t-T_0)/P$.
 All these quantities are known when the spectroscopic and
interferometric constraints are combined, except the ratio $(M_1+M_2)/M_2$.
Since the primary is a post-AGB star, its mass can only take a narrow range of values (typically 0.5-0.9\,M$_\odot$) as the future WD is only surrounded by the thin remaining envelope.
We therefore use the measured spectroscopic mass function:
\begin{equation}
    f(m) = \frac{M^3_2\sin^3(i)}{(M_1+M_2)^2} \rightarrow M_1 \propto M_2
\end{equation}
to estimate a range for $(M_1+M_2)/M_2$.
Since the mass ratio is used in the interferometric fit to fix the
position of the center of mass, we repeat the fit with different mass ratio values. 
We find the mass ratio to have a small influence on the fit, but we take a conservative approach and continue our analysis with the 2-$\sigma$ upper limit on the fitted binary angular separation (0.91\,mas; as well as on the inclination: 23$^\circ$). 
The choice for an upper limit is also motivated by the fact that the best-fit separation (0.81 mas) from Paper~I is significantly below the formal resolution limit of the observations ($\lambda/2B_\mathrm{max}$=1.25\,mas). 
One expects in this case a degeneracy between the best-fit separation and the flux contribution from the secondary \citep[e.g.][]{Willson2016} that was indeed observed in Paper~I. 
Hence we derive 2-$\sigma$ lower bounds on the distance and luminosity of IRAS08544-4431, as listed in Table\,\ref{tab:fundamentalparam} and that we will use in the rest of the paper. 
The mass of the central object is estimated from the relation between the (core) mass and the luminosity of a post-AGB star \citep{Vassiliadis1994}:
\begin{equation}
    \frac{L_1}{L_\odot} = 56694 (\frac{M_1}{M_\odot}-0.5) \rightarrow L_1 \propto M_1 ,
\end{equation}
and the measured mass function. 
This results in a total mass of $M_\mathrm{tot} = M_1+M_2 \sim 2.4\,M_\mathrm{\odot}$.
This mass is compatible with the analysis of the ALMA maps of $^{12}$CO and $^{13}$CO $J=3-2$ lines that trace Keplerian motion of the circumbinary disk \citep{Bujarrabal2018}.

 \subsection{Determination of fundamental stellar parameters}

\begin{table}[!t]
\caption{Fundamental parameters of the primary of IRAS08544-4431.}
\begin{center}
\begin{tabular}{l|c}
Parameter & Value  \\
 \hline
 Effective temperature $T_\mathrm{eff}$ (K) & 7250 (fixed)\\
 Metallicity $[Fe/H]$ (dex) & -0.5 (fixed) \\
 Angular diameter $\theta$ (mas) & 0.48 $\pm$ 0.08\\
 Reddening $E(B-V)$ (mag) & 1.35 $\pm$ 0.12
\end{tabular}
%\tablefoot{}
\end{center}
\label{tab:fundamentalparam2}
\end{table}%

\begin{figure}[!t]
    \centering
    \includegraphics[width=10cm]{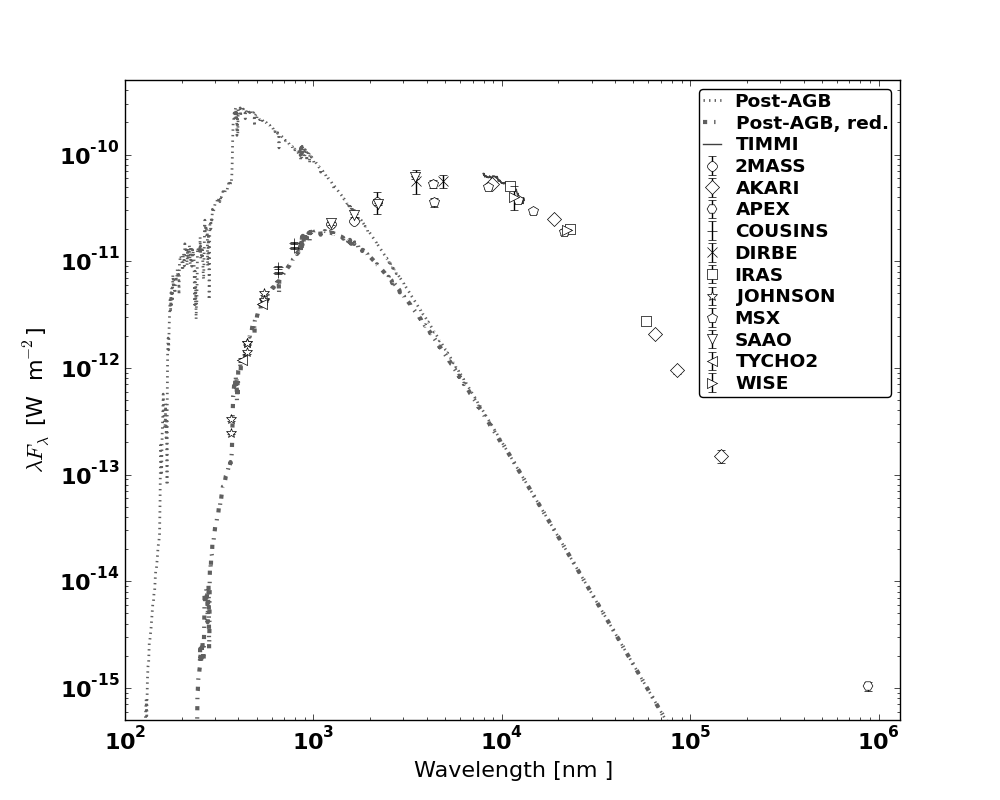}
    \caption{SED of IRAS08544-4431. The grey dotted and dash-dotted lines represent the primary's original and reddened photospheric spectrum, respectively.}
    \label{fig:SED}
\end{figure}

Because no direct flux contribution has been detected from the companion in the high-resolution optical spectra \citep{2003aamaas}, we adopt a single-star atmosphere model to fit the stellar part of the SED (\textless1.5\,$\mu$m). 
Assuming spectroscopically-derived photospheric parameters \citep{2003aamaas} we fit ATLAS models \citep{Castelli2003} to the SED to derive the angular diameter of the primary and the total line-of-sight reddening to the system \citep{2004aspcfitzpatrick,2013aahillen}. 
To improve the accuracy of the fit, we also include the stellar part of the H-band flux as determined in Paper~I (59.7\% of the total flux). 
We use a grid-based method and a $\chi^2$ statistic for the parameter estimation \citep{2004aspcfitzpatrick}. 
The stellar parameters of the best-fit model are listed in Table\,\ref{tab:fundamentalparam2}.

\subsection{Radiative transfer model}

\begin{table*}[!t]
\caption{MCMax radiative transfer model parameters for IRAS08544-4431.}
\begin{center}
\begin{tabular}{l|c|r|c}
Model parameter & Grid range & Best-fit value & Acceptable range \\
 \hline
 $R_\mathrm{in}$ (au) & 5 - 12 & 8.25 & [8; 9.25]\\
 $R_\mathrm{mid}$ ($R_\mathrm{in}$) & [2.0; 2.5; 3.0; 3.5; 4.0] & 3.0 & [2.5; 3.0] \\
 $R_\mathrm{out}$ (au) & 200 (fixed) & 200 & - \\
 $p_\mathrm{in}$ & -1.0; -1.5; -2.0; -2.5; -3.0 & -1.5 & [-2; -1] \\
 $p_\mathrm{out}$ & 1.0 (fixed) & 1.0 & -\\
 $a_\mathrm{min}$ ($\mu$m) & 0.01; 0.10; 1.00; 3.00 & 0.10 & [0.1; 3]\\
 $a_\mathrm{max}$ (mm) & 1.0 (fixed) & 1.0 & - \\
 $q$ & 2.75; 3.00; 3.25; 3.50 & 2.75 & [2.75; 3.25]\\
 Gas/dust ratio & 100 (fixed) & 100 & -\\
 $\alpha$ & 0.001; 0.01 & 0.01 & [0.001; 0.01]\\
 $M_\mathrm{dust}$ (M$_\odot$) & 0.0001 - 0.01 & 0.002 & [0.002] \\
 \hline
 \hline
 $\chi^2_\mathrm{red,V^2}$ & - & 2.4 & [2.4; 3.4]\\
 $\chi^2_\mathrm{red,CP}$ & - & 41 & [33; 65] \\
 $\chi^2_\mathrm{red,SED}$ & - & 17.9 & [14.7; 55.7] \\
\end{tabular}
%\tablefoot{}
\end{center}
\label{tab:SED}
\end{table*}%

\subsubsection{The radiative transfer code: MCMax}

We want to reproduce our photometric and interferometric observations with a self-consistent physical model of the dusty circumbinary disk, to investigate the physical conditions at the inner rim. 
We assume the disk is in hydrostatic equilibrium and that its energetics are fully determined by the stellar irradiation. 
The code that we use to compute disk structures, MCMax \citep{2009AAMin}, is based on the Monte Carlo method. 
Photon packages are randomly emitted by a source at the origin of the coordinate system, which are then absorbed or scattered by dust that is distributed in an axisymmetric geometry. 
The thermal structure is hence determined from the interaction of the dust with the stellar radiation (i.e., a passively heated disk in which the gas is in thermal equilibrium with the dust). 
The radial distribution of the gas and dust is a basic input of the model, but the vertical structure is obtained by solving the equation of hydrostatic equilibrium (i.e., the vertical component of the local gravitational force is balanced by the local gas pressure gradient). 
The temperature and density profiles in the disk are iteratively determined. 
The Monte Carlo process is very efficient to explore a reasonable parameter space of disk and dust properties and to compute a collection of observables when the disk is axisymmetric with a single central heating source.
Therefore, the model takes into account the binary in terms of the central mass but sets a gravitational potential of a disk-centred single star, ignoring any gravitational perturbations produced by the binarity.
We show that our observables, which probe the thermal and density structure of the innermost circumbinary material, are well reproduced with an axisymmetric disk model in hydrostatic equilibrium. 
We also keep this limitation in mind for further discussion.

We summarize the main properties of our radiative transfer models:
\begin{itemize}
    \item The vertical extension of the disk is set by the hydrostatic equilibrium,
    \item photon scattering by the dust is included in a full angle-dependent way,
    \item the composition of the dust is assumed to be an ISM-like mixture of silicates in the DHS (Distribution of Hollow Spheres) approximation \citep{2007aamin},
    \item the size of the dust grains follows a power-law distribution,
    \item a grain-size-dependent settling of dust, counteracted by turbulence, is included self-consistently \citep{2012aamulders},
    \item we adopt a double-power-law formalism to parameterize the surface density distribution \citep{2015aahillen}.
\end{itemize}

These properties translate into the following parameters. 
First, the double-power-law surface density distribution is parameterized with five parameters: the inner and outer radius ($R_\mathrm{in}$ and $R_\mathrm{out}$), the turnover radius ($R_\mathrm{mid}$), the power-law exponent in the inner part and in the outer part ($p_\mathrm{in}$ and $p_\mathrm{out}$). 
The grain size distribution requires three parameters: the minimum grain size ($a_\mathrm{min}$), the maximum grain size ($a_\mathrm{max}$) and the grain size power law exponent ($q$). 
There is also a measure of the total dust mass ($M_\mathrm{dust}$) and the total gas mass (for the vertical structure computation; here in the form of a global gas/dust ratio). 
Finally, the turbulence is parameterized with the $\alpha$-prescription \citep{Shakura1973}, in which $\alpha$ is a scale parameter called the turbulent mixing strength \citep{2012aamulders}. 
We computed a grid of models and we refer to Table\,\ref{tab:SED} for the ranges of the different parameters of our grid.

\subsubsection{Goodness-of-fit criteria}

We evaluated the quality of our models by comparing the SED and the PIONIER squared visibilities with the corresponding synthetic observables. 
As our model does not include the binary properties, we only try to match the \modif{radial morphology} of the detected near-IR emission in all channels, and not its detailed azimuthal profile.
We therefore do not include the closure phase residuals in our merit function because none of our models can successfully reproduce the detected asymmetry. 
This is not surprising because the structure of the radiative transfer models is axisymmetric.
At moderate or high disk inclination the radiative transfer model can produce an asymmetric image, due to the disk self-absorption effects, and hence non-zero closure phases.
However, in our case the disk is seen almost pole-on and these effects are negligible.
We tested this by computing models with various inclinations, finding a preference for 20$\pm$7.5$^\circ$. 
On the other hand, our tests do indicate a difference with respect to the disk position angle that was derived with the parametric model in Paper~I (by about -30$\pm$15$^\circ$).

To fit the interferometric data we need to reproduce the integrated flux of each component (stars, disk) simultaneously with the scale from which this flux arises. 
The squared visibilities are very sensitive to the relative flux fractions. 
Here, we fix the stellar flux contributions on the basis of the values derived from the parametric model of Paper~I. 
Similarly, the location of the companion, its temperature, and its flux contribution are fixed in these models.
In other words, we neglect the influence of the duplicity of the central heating source in the computation of the disk structure (the luminosity of the secondary is negligible compared to the primary). 
However, in Paper~I we detect 6\% of the flux in $H$-band is coming from the position of the secondary. 
The binary has therefore a direct influence in the interferometric data and the SED in the $H$-band. 
We summarize the practical requirements for a model to be considered good: it needs to 1) contribute about 21\% of the total flux at 1.65 micron in the form of thermal emission from the inner rim of the disk, as well as 15\% in the form of scattering on larger scales, 2) fit the position and shape of the visibility minimum and secondary maximum, and 3) be in good agreement with the full SED.

\begin{figure*}[th]
    \centering
    \includegraphics[width=20cm]{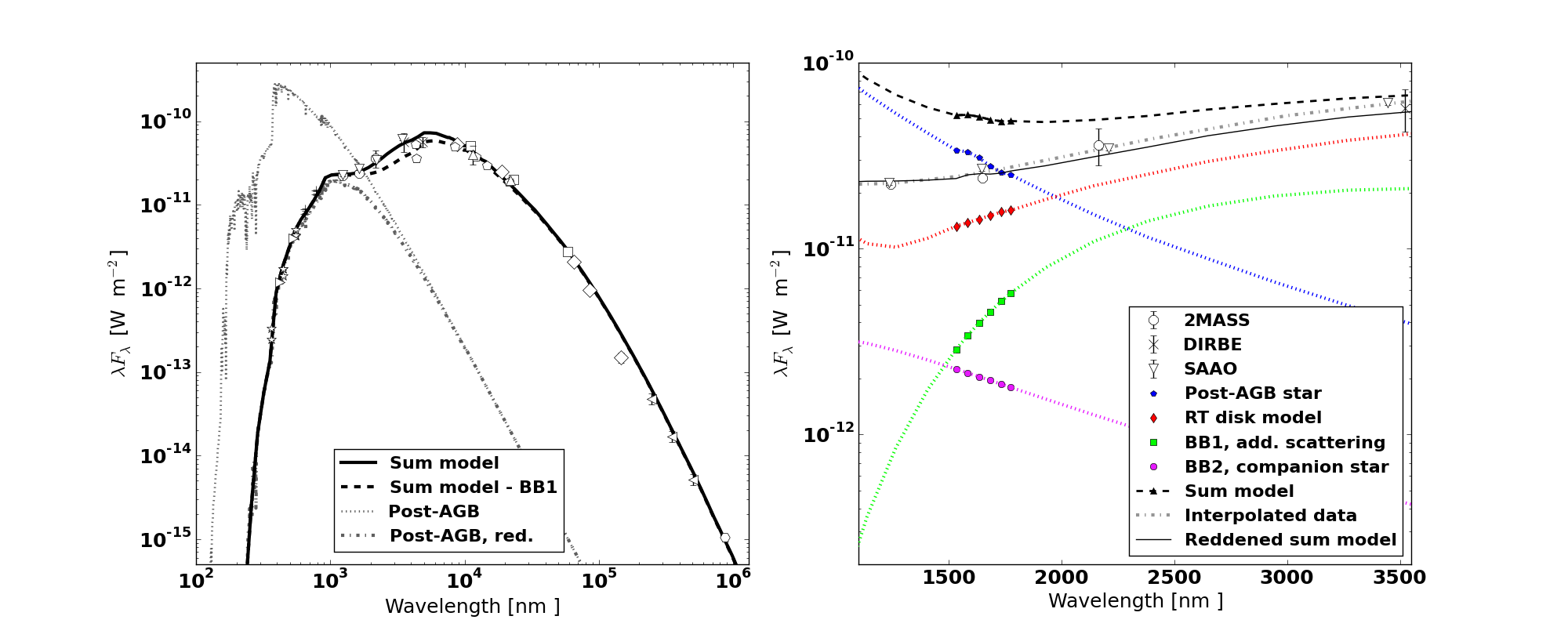}
    \caption{\modif{Comparison of the best fit model photometry to the SED.} Left: the SED of IRAS08544-4431. The grey dotted and dash-dotted lines represent the primary's original and reddened photospheric spectrum, respectively. Right: A zoom-in on the near-infrared wavelengths.}
    \label{fig:SEDfit}
\end{figure*}

\begin{figure}[th]
    \centering
    \includegraphics[width=9.5cm]{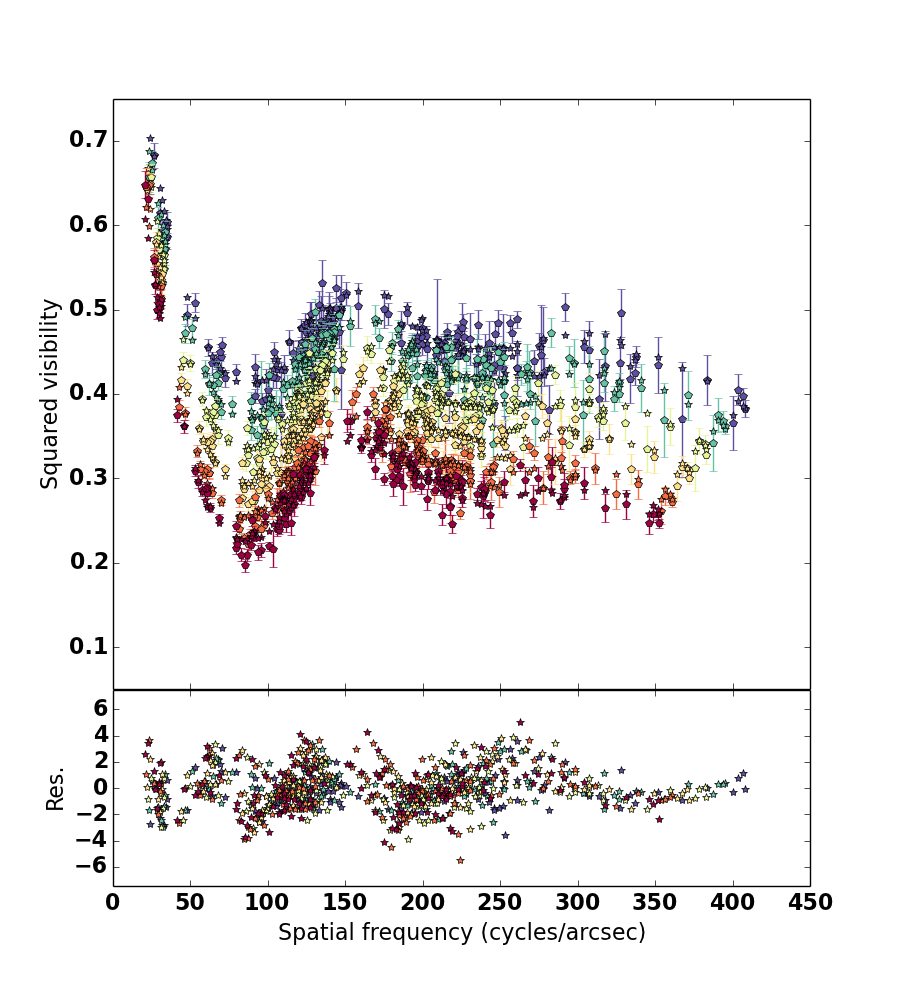}
    \caption{\modif{Comparison of the best fit model squared visibities to the PIONIER dataset.} Top: the squared visibilities form the best fit radiative transfer model (stars) and data squared visibilities (pentagons) versus the spatial frequency. The color represents the wavelength (blue: 1.55$\mu$m; red: 1.8$\mu$m). Bottom: the residuals normalized by the error.}
    \label{fig:V2fit}
\end{figure}

\begin{figure}[th]
    \centering
    \includegraphics[width=9.5cm]{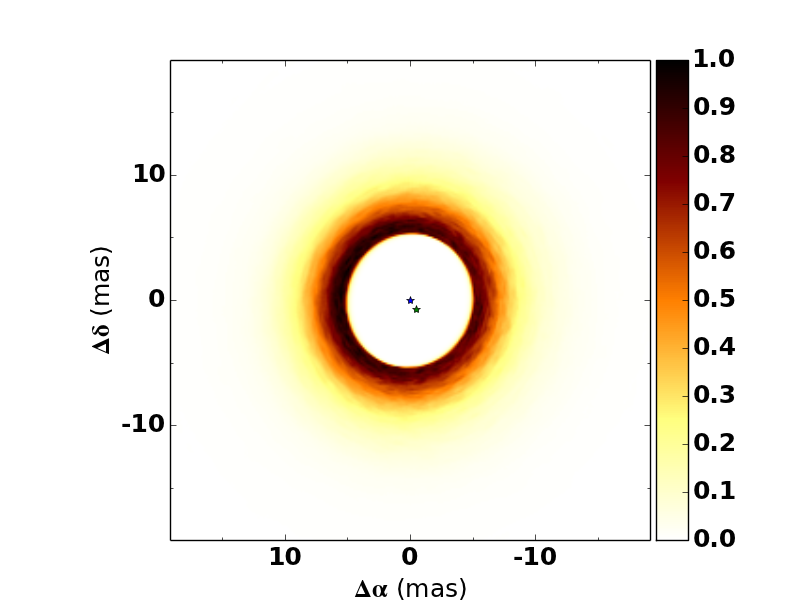}
    \caption{Image of the best fit radiative transfer model of IRAS08544-4431. The primary star is in blue and the secondary is in green.}
    \label{fig:RTimg}
\end{figure}
  
\subsubsection{An issue with the over-resoled flux seen by PIONIER}

Most models that match the mid- to far-IR SED produce too much thermal emission in the near-IR, because the inner rim is too hot and optically thick. 
The models that are thermally in the right range seem to underestimate the $\sim$15\% over-resolved flux (typically 5-10\% is reached). 
This discrepancy also has a color: more flux is lacking at long- than at short wavelengths. 
We estimate color temperatures for the over-resolved flux of $\sim$3500 K, which is well above the 2400\,K detected in the real data. 
To decide on the best-fit model, we select a subset of models among our grid in which the total 1.65\,$\mu$m disk flux does not exceed the 36\% derived parametrically (i.e., smaller than the sum of the ring and background flux). 
For each of these MCMax models, we arbitrarily add over-resolved flux (i.e., background flux like in the parametric case), until the sum of all flux components reaches 100\% at 1.65\,$\mu$m. 
The temperature of this added flux is determined by minimizing the gradient with respect to wavelength of the average of the visibility residuals. 
Finally, we select the model with the lowest visibility reduced chi-square ($\chi^2_\mathrm{r}$).

\subsubsection{The best-fit model}

The parameter values of our best model are listed in Table\,\ref{tab:SED}. 
For this model an additional component of 8.1\% is needed at 1.65\,$\mu$m, and with a color temperature of 1000$\pm$250 K. 
The synthetic visibilities are shown in Fig.\,\ref{fig:V2fit} along with the residuals.
The SED of our best-fit model is displayed in Fig.\,\ref{fig:SEDfit}, along with an enlarged view of the near-IR wavelength region in order to highlight the various constituents. 
Our model matches the observed fluxes from sub-$\mu$m to sub-mm wavelengths. 
The pure MCMax+binary model (i.e., without the additional over-resolved flux) is shown as well. 
A simple black-body for the additional component extrapolates correctly over the whole near-IR wavelength range, but overestimates the observations slightly between 4 and 8\,$\mu$m.
  
In order to estimate the acceptable range of the model parameters we selected the acceptable models that have $\chi_\mathrm{red, V^2}^2 < \mathrm{min} (\chi_\mathrm{red, V^2}^2) + 1$.
The range of values of these parameters is shown in Table\,\ref{tab:SED}.
We find that $R_\mathrm{in}$ as well as $R_\mathrm{mid}$ is well constrained by the interferometric dataset which is not surprising as the spatially resolved data strongly constrain the disk structure.
We also note that the minimum size of a dust particle ($a_\mathrm{min}$) can only be larger that the one of the best-fit model.
The turbulent mixing length ($\alpha$) is not well constrained by the interferometric dataset.
Finally, M$_\mathrm{dust}$ is well constrained by the model. 
However, the models do not match the extended flux and additional dust would possibly be needed.
The good constrain\modif{t} on the dust mass is therefore not exempt of a systematic error.

\subsection{Image reconstruction comparison}

To compare the radiative transfer model with our data, we do an image reconstruction on the synthetic data built from the best-fit radiative transfer model.
To be able to compare the reconstruction of the image on the original dataset to the one we do on the radiative transfer model, we use the same image reconstruction approach \citep[SPARCO;][]{2014aakluska} as in Paper~I with the same image configuration (number and angular size of the pixels), regularization (type, weight) and parametric model parameters (flux of the primary, of the secondary, the positions of the secondary, $x_\mathrm{sec}$ and $y_\mathrm{sec}$, and the spectral index of the environment, \denv).
These parameters are summarized in Table\,\ref{tab:imgrec}.

\begin{table}[!t]
\caption{Parameters used in the image reconstruction.}
\begin{center}
\begin{tabular}{l|c}
Parameter & Value  \\
 \hline
Number of pixels & 512$\times$512\\
Pixels size \modif{(mas)} & 0.15\\
 \hline
Regularization & Quadratic smoothness\\
Weight & 10$^{12}$\\
\hline
Diameter of the primary \modif{(mas)}  & 0.5 \\
Flux of the primary \modif{(\%)} & 59.7 \\
Flux of the secondary \modif{(\%)} & 3.9 \\
$x_\mathrm{sec}$ \modif{(mas)} & -0.44 \\
$y_\mathrm{sec}$ \modif{(mas)} & -0.68 \\
\denv & 0.42 
\end{tabular}
%\tablefoot{}
\end{center}
\label{tab:imgrec}
\end{table}%

\begin{figure*}[th]
    \centering
    \includegraphics[height=6cm]{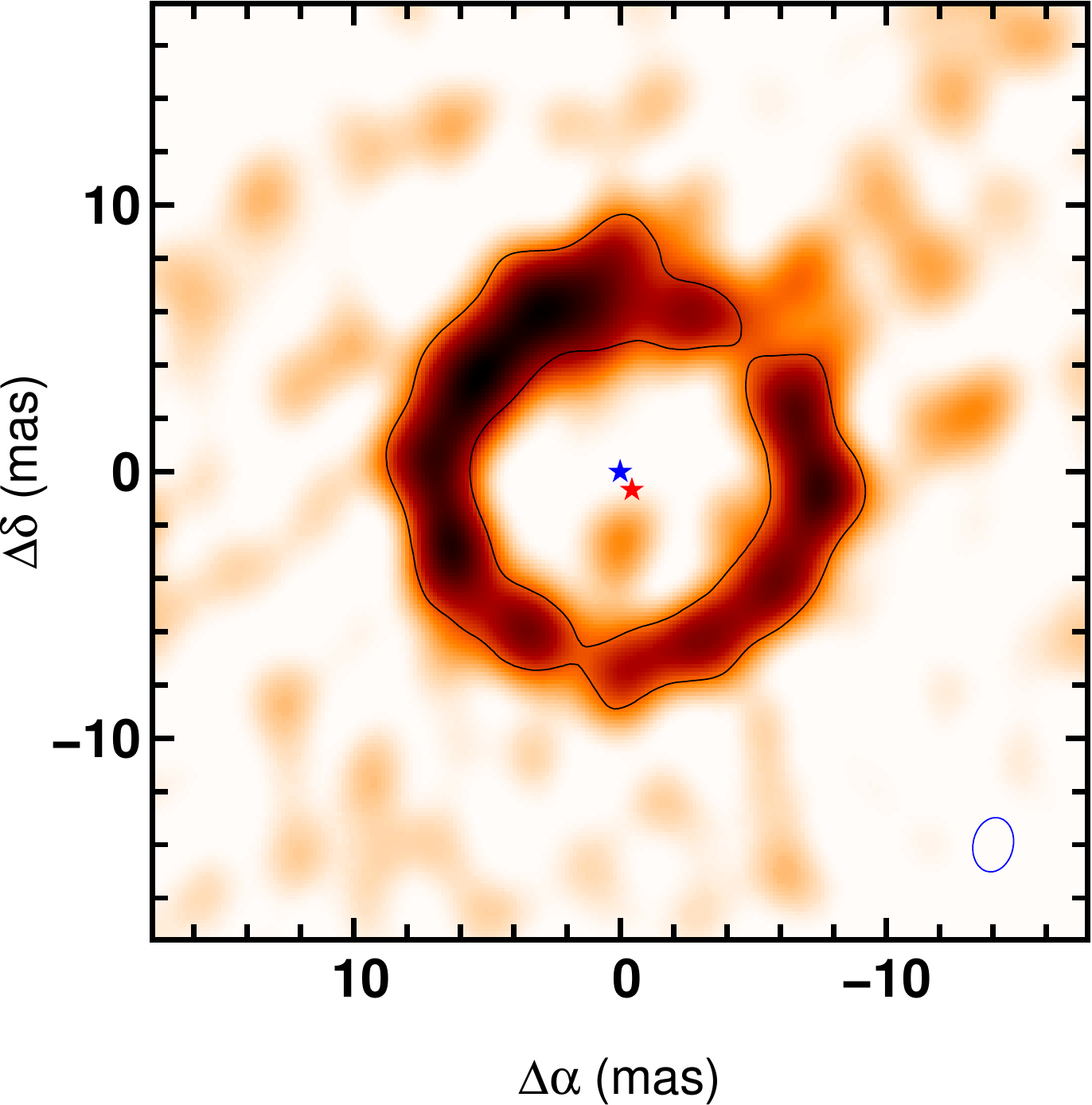}
    \includegraphics[height=6cm]{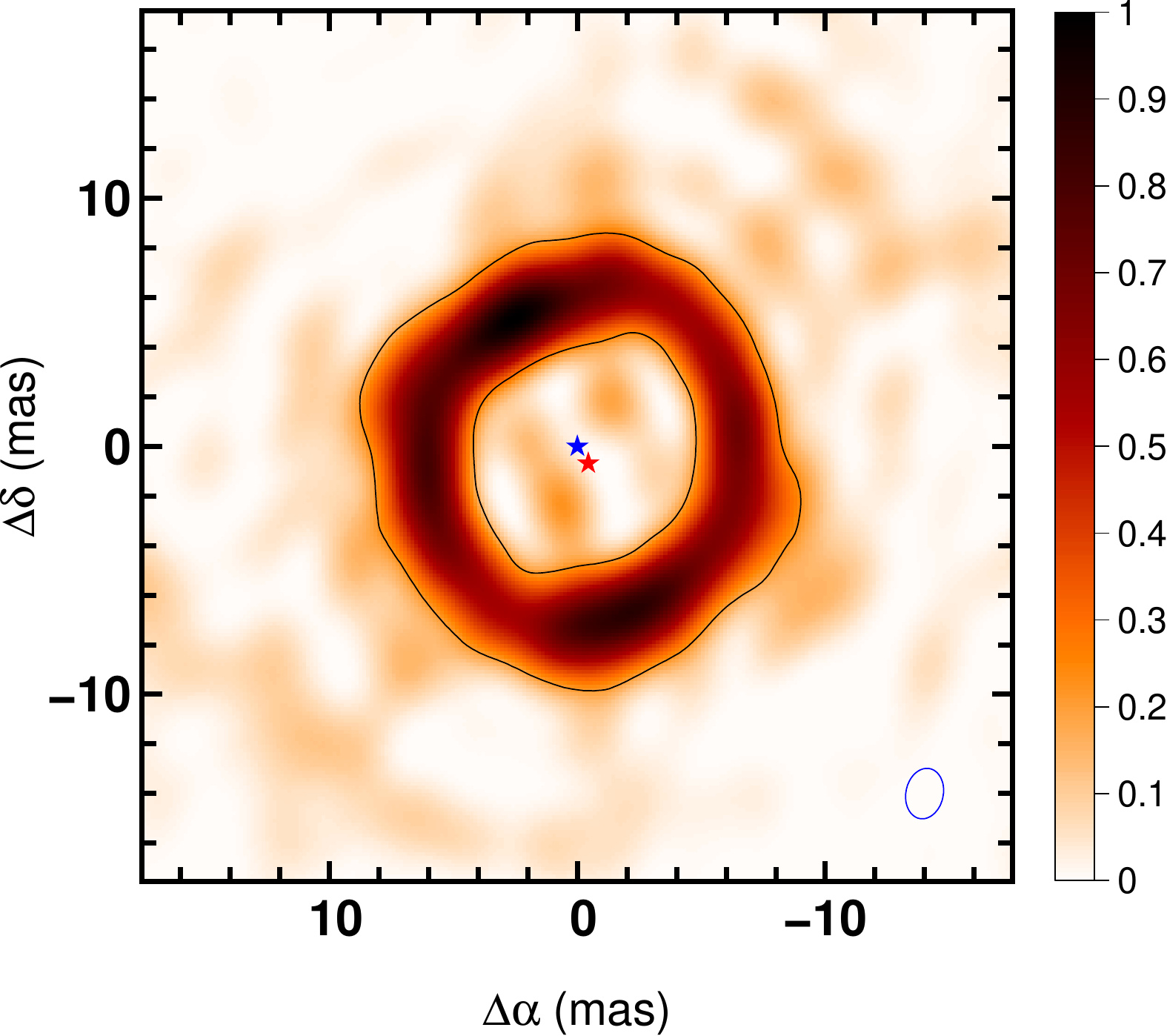}
    \caption{Image reconstructions on the actual dataset of IRAS08544-4431 (left) and on the best-fit radiative transfer model (right). The blue star indicate the position of the primary star and the red star of the secondary one. The solid line contour is the 5-$\sigma$ significance contour.}
    \label{fig:RTimgrec}
\end{figure*}

The two images are similar (see Fig.\,\ref{fig:RTimgrec}).
The rings have the same radius and are very similar in brightness.
This is not surprising as the radiative transfer model was selected because it is matching the squared visibilities.
The flux located outside the ring is also similar in both images: it is the over-resolved flux and its distribution is mainly ruled by the \uv-coverage.
However, the ring azimuthal brightness distribution, which is coded in the closure phases, looks different in both images.

\begin{figure*}[th]
    \centering
    \includegraphics[height=7cm]{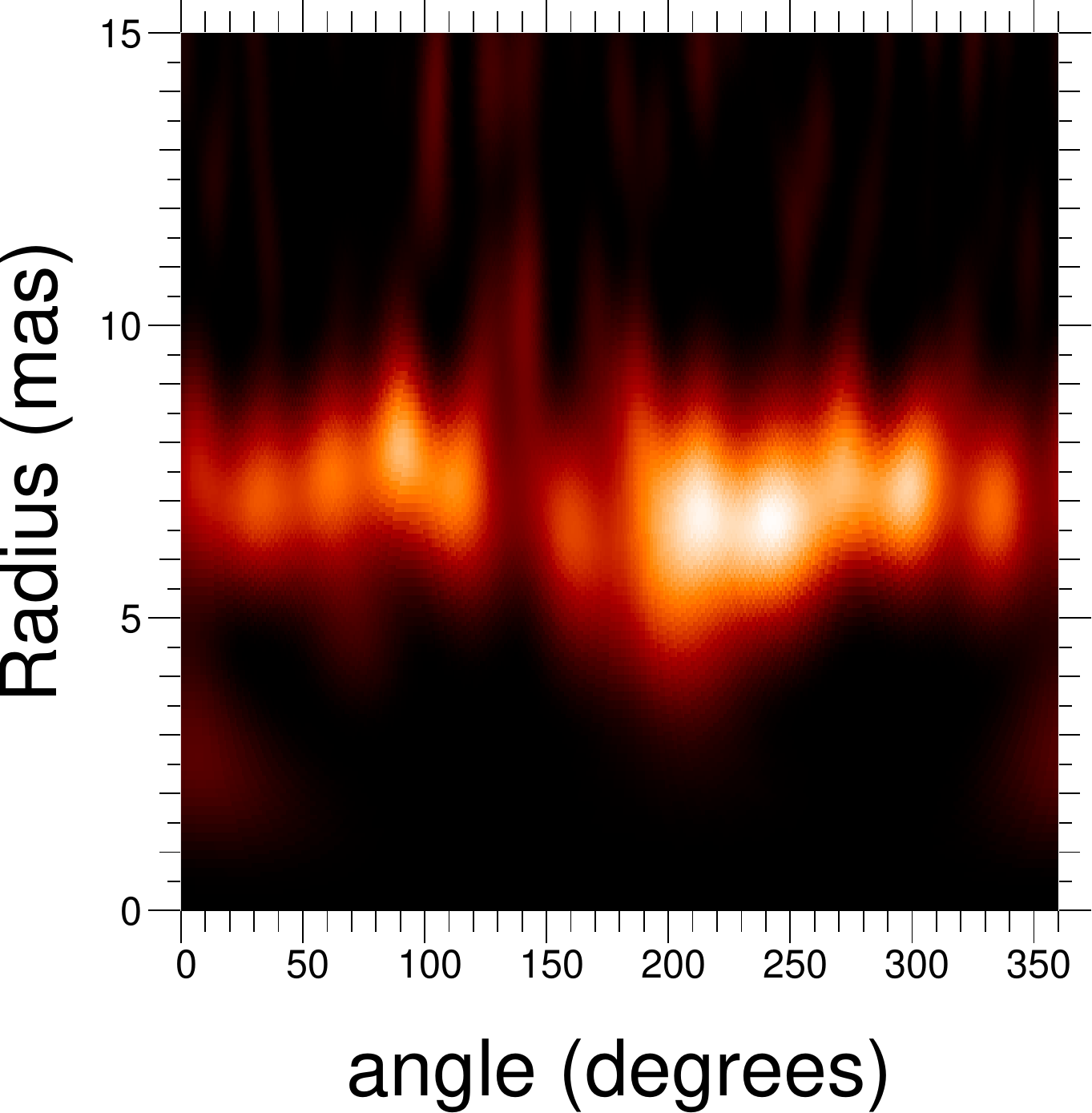}
    \includegraphics[height=7cm]{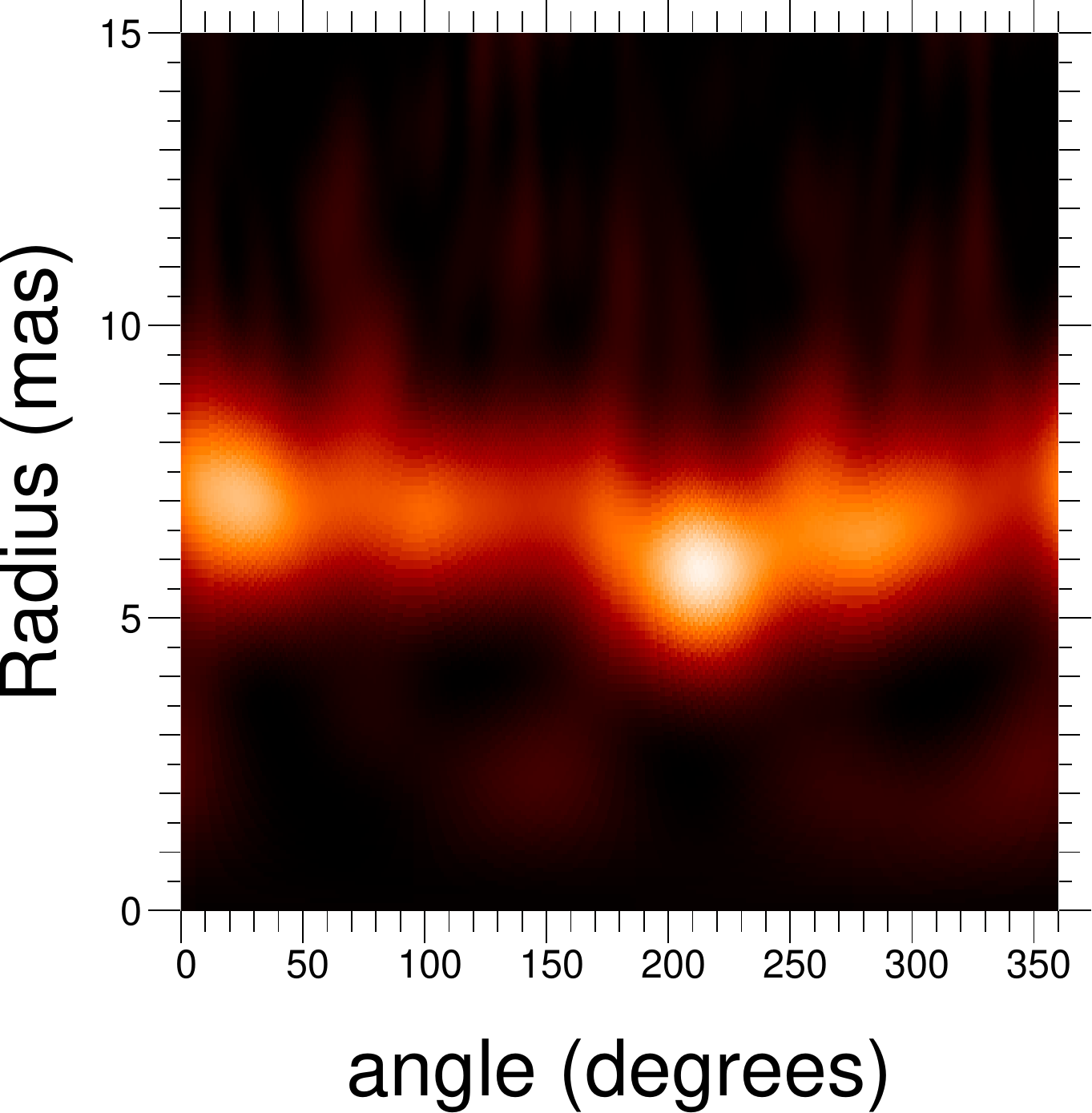}
    \caption{Polar plot of the image reconstructions on the dataset (left) and on the best fit model (right).
    }
    \label{fig:polar}
\end{figure*}

We investigate the rim azimuthal brightness distribution by making polar plots (see Fig.\,\ref{fig:polar}).
The polar plots are corrected for the disk orientations ($i$=19$^\circ$; $\theta$=6$^\circ$).
We can see that both images show that the ring emission is centred at a radius of $\sim$7\,mas.
Whereas the ring from the radiative transfer model looks smooth and continuous, the actual ring shows azimuthal discontinuities (for example at 140$^\circ$ or 355$^\circ$).
For both polar plots the light seems to be closer to the centre at an angle of $\sim$210$^\circ$.
The image on the actual data is also displaying variations in width and dips that are not seen in the image reconstruction of the radiative transfer model.

\begin{figure}[th]
    \centering
    \includegraphics[width=5cm]{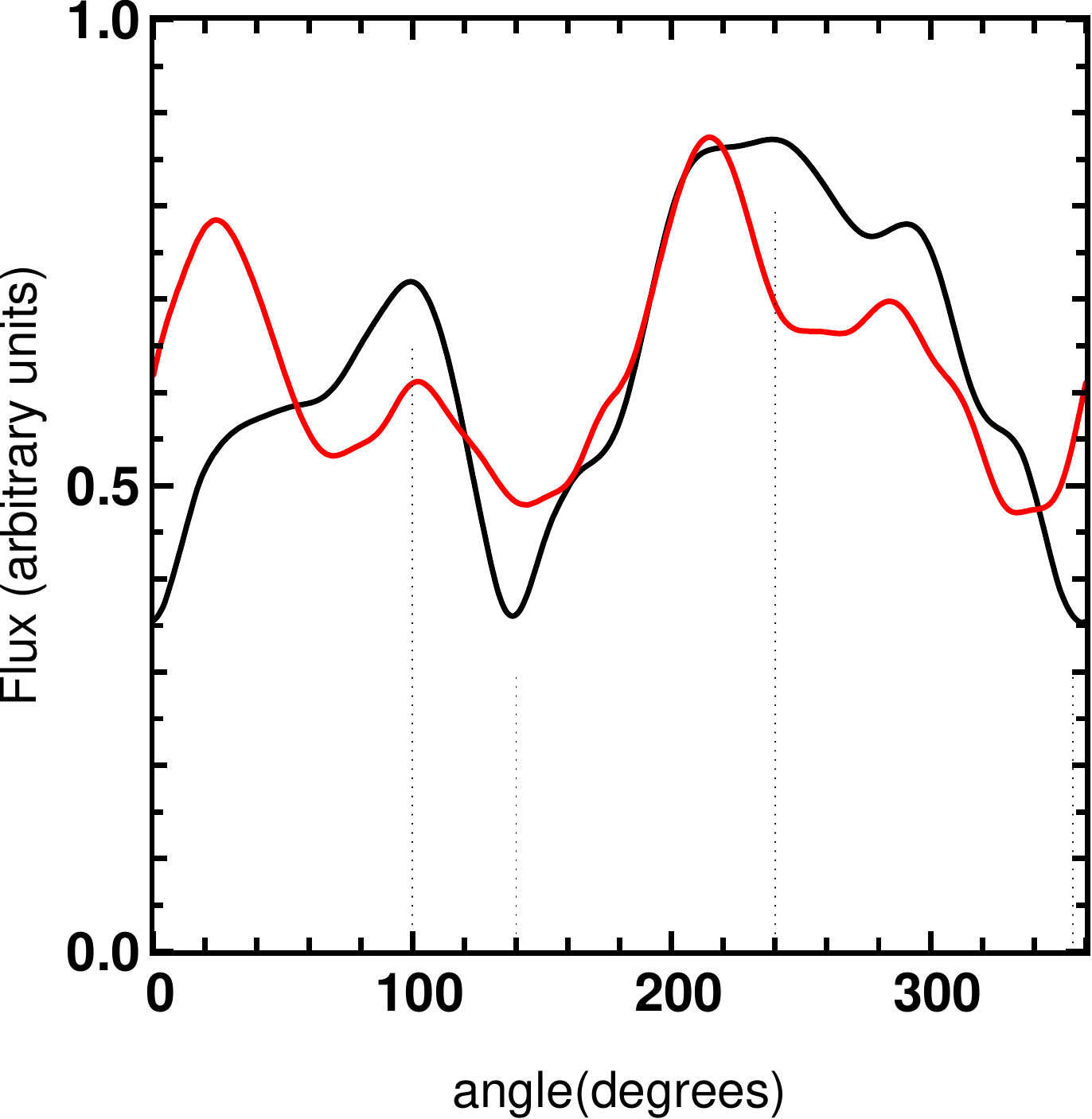}
    \caption{Azimuthal profile of the image reconstruction of the inner rim (between 5 and 9\,mas from the central star) of IRAS08544-4431 (black) and of the best-fit radiative transfer model (red).
    }
    \label{fig:azim}
\end{figure}

\begin{figure}[th]
    \centering
    \includegraphics[width=5cm]{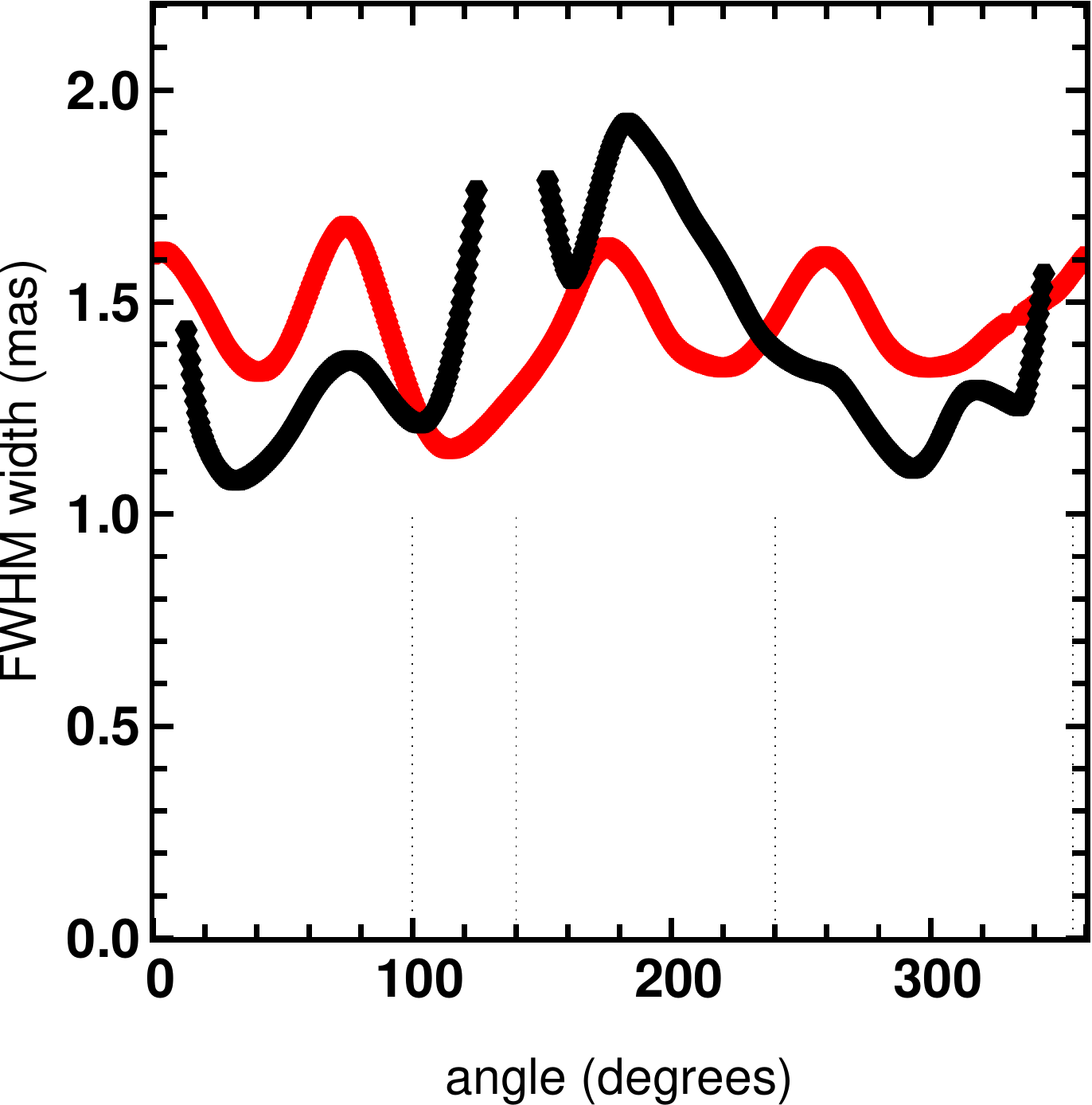}
    \caption{Width of the Gaussian fitted in function of the position angle. Black: Image of IRAS08544-4431; red: image of the best fit radiative transfer model.
    }
    \label{fig:width}
\end{figure}

To investigate the rim azimuthal brightness variations more quantitatively we summed up the flux between 5 and 9\,mas from the centre for 80 azimuthal segments (see Fig.\,\ref{fig:azim}).
The profile from the actual data on IRAS08544-4431 displays indeed more variations than the one on the radiative transfer model.
The change in the brightness throughout the ring is also larger in the actual disk having a ratio between maximum and minimum of 2.5 where the radiative transfer model image has a ratio of 1.8.
Also the relative peak-to-peak variations is larger for the actual image (82\%) than on the model (59\%).

We report the maximas (at 100$^\circ$ and 240$^\circ$) and the minimas (or dips, at 140$^\circ$ and 355$^\circ$) from the actual image on Fig\,\ref{fig:azim}.
These extremas are not found in the radiative transfer model image at the same angles. This suggests strong azimuthal variations in the intrinsic object morphology.

Lastly, we fit 1D-Gaussians to the polar plot of both images to retrieve the width of the inner rim emission in function of the azimuthal angle.
On Fig.\,\ref{fig:width} we can see the width variations with azimuth.
The width corresponding to low fluxes (half of the maximum flux, corresponding to the rings dips) are hidden as the width has no astrophysical meaning.
The variations in the width of the actual disk rim are larger (peak-to-peak of 0.85\,mas) than in the one of the radiative transfer model (peak-to-peak of 0.53\,mas).
We remark that in the largest flux maximum (at 240$^\circ$) the width is particularly large : 2.00$\pm$0.02\,mas.

 \section{Discussion}
 \label{sec:dis}

 \subsection{On the best fit RT model}

The excellent agreement between our best model and the resolved observations gives strong support to our physical interpretation of the circumbinary material in this object: its structure is \modif{well reproduced by our model} of a dusty settled disk.
The best-fit grain size distribution shows that grain growth is significant. 
Sub-micron-sized particles hardly contribute to the total opacity in this disk, while the mm-sized grains are an important source of opacity and explain the sub-mm fluxes.
Our double power-law radiative transfer model fits well both the photometric and interferometric data.
Significant grain-growth is a common property of the disks around post-AGB stars \citep{deruyter05, gielen11, 2014aahillen, 2015aahillen} and can be used as a tracer of longevity. 

\begin{figure}[!t]
    \centering
    \includegraphics[width=6cm]{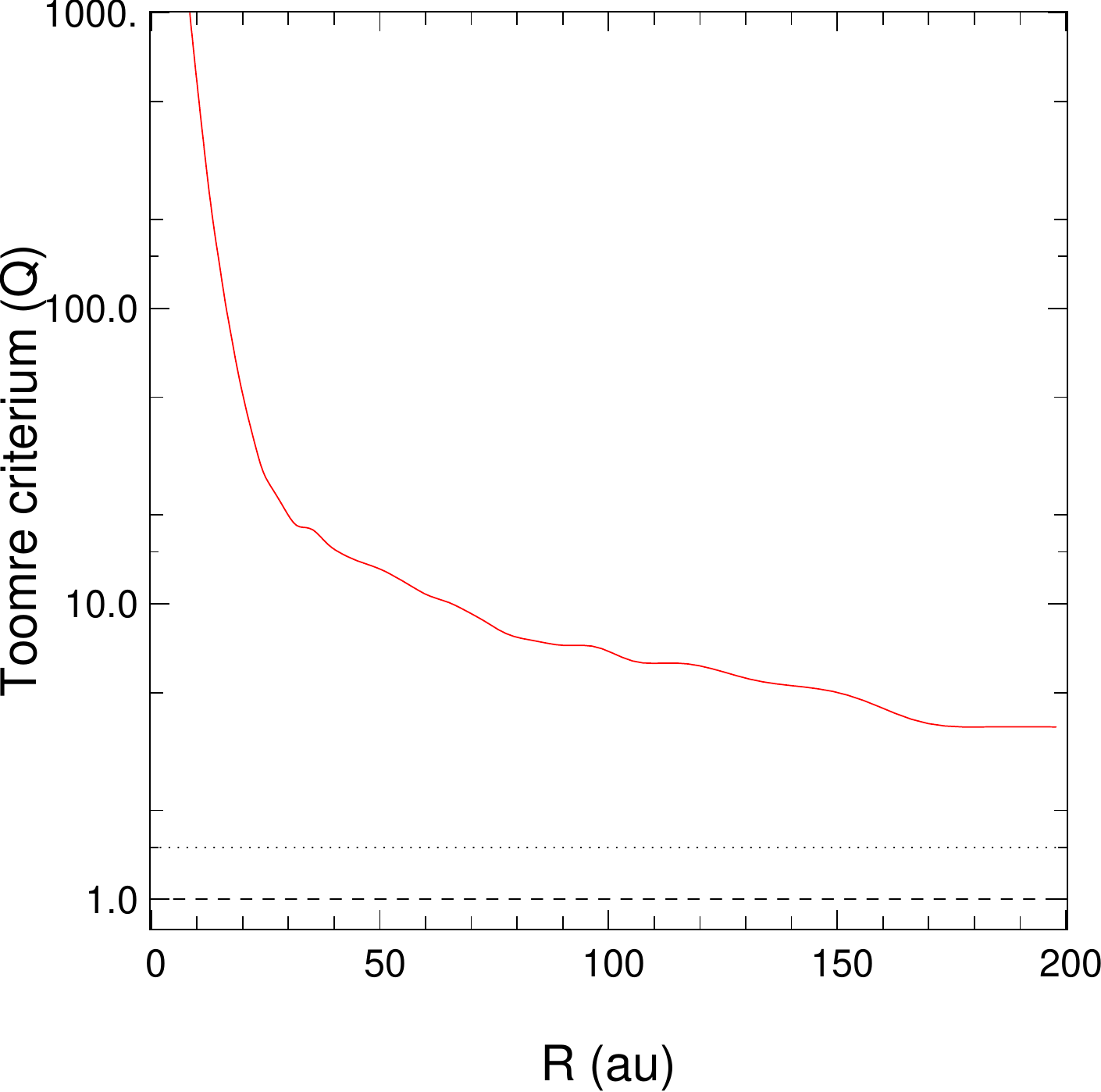}
    \caption{\modif{Toomre criterium ($Q$) for the best-fit disk model against the disk radius.
    The dashed line represent the $Q$=1 and the dotted line $Q$=1.5.}
    }
    \label{fig:Toomre}
\end{figure}

\modif{The best fit model is gravitationally stable as the Toomre criterium stays above 1 for axi-symmetric disks or 1.5 for disks with asymmetric structures \citep[see Fig.\,\ref{fig:Toomre};][]{Toomre1964,Papaloizou1991,Mayer2004,Durisen2007}.
This reinforces our choice of modeling the disk in hydrostatic equilibrium. 
As the disk is more likely to be gravitationally unstable in the outer parts, the notion of disk stability can be estimated more reliably by combining optical/near-IR scattered-light imaging with resolved data of the (sub-)mm continuum as well as of the gas. 
This will provide better constraints on the disk extension, mass, scale-height, and stratification.
}

A detailed full model of the gas and the dust is beyond the scope of this paper. Such a model of the system needs to include, from wide to narrow, the slow gaseous outflow, the sub-Keplerian and Keplerian part, the inner puffed-up rim, the circum-companion accretion disk as the origin of the fast outflow and the evolved luminous primary.

 \subsection{Inner rim compatible with the dust sublimation radius}
 
The best fit inner rim radius is 8.25\,au.
Several processes can set the radius of the inner disk rim.
One of them can be the dynamical disk truncation due to the inner binary.
The instantaneous separation of the inner binary is about 1.2\,au.
With a mass ratio of $\sim$0.45 the ratio between the binary semi-major-axis and the disk gap radius is $\sim$1.7 \citep{Artymowicz94}.
In our case the disk truncation radius would then correspond to $\sim$2\,au.
Unless there is a third component in the system the dynamical truncation is not determining the location of the disk inner rim.

A more likely possibility is that this rim corresponds to the dust sublimation radius.
The disks inner rim around young stellar objects (YSOs) are known to be ruled by dust sublimation physics and therefore by the luminosity of the central star.
Therefore their sizes are proportional to the square-root of the luminosity of the central star \citep{Monnier2002,Lazareff2017} following this equation:
\begin{equation}
    R_\mathrm{rim} = 1.1 (C_\mathrm{bw}/\epsilon)^\frac{1}{2}\Bigg(\frac{L_*}{1000L_\odot}\Bigg)^\frac{1}{2} \Bigg(\frac{T_\mathrm{sub}}{1500 \mathrm{K}}\Bigg)^{-2},
\end{equation}
where $R_\mathrm{rim}$ is the rim radius, $C_\mathrm{bw}$ is the backwarming coefficient \citep[see][for more details on this coefficient]{Kama2009}, $\epsilon$ is the dust grain cooling efficiency, $L_\mathrm{*}$ is the stellar luminosity and $T_\mathrm{sub}$ is the dust sublimation temperature.
Applying this equation to IRAS08544-4431 we find a dust sublimation radius of 8.2\,au assuming classical values for inner rims of disks around YSOs ($T_\mathrm{sub}$=1500\,K, $C_\mathrm{bw}$=4, $\epsilon$=1).
This value is remarkably close to the one we find in our observations.
We can therefore conclude that the inner rim is indeed likely ruled by dust sublimation due to the heating from the central star.

\subsection{Origin of the asymmetry}

In this paper we show that the inner disk rim around IRAS08544-4431 cannot be considered as fully axisymmetric.
The inner binary is very likely to be at the origin of this asymmetry.
Here we discuss the possible mechanisms that can make the inner binary disturb the inner disk rim.
A dust grain located at the inner disk rim receives a variable illumination from the central star and is in a variable gravitational field during an orbital period.

The star illuminating the disk is also the less massive one in the binary system.
This means that for a full orbital period, a given part of the disk will see the primary star at different distances.
The distance can vary between 7.0\,au and 9.5\,au.
If we compute the averaged distance over one orbital period it will differ from the distance to a fixed star at the centre of the disk by less than 0.04\,au.
It will change the effective luminosity perceived by a point in the disk by less than a 1\%.
At the closest and furthest points, however, the stellar flux seen by a dust grain at the inner rim varies by up to 18\%.
If the radiative timescale is shorter than the orbital period, this, alone, can already generate a disk scale-height perturbation because of hydro-static equilibrium.

Let us compute the typical disk scale-height by computing the hydrostatic equilibrium for extreme phases of the binary w.r.t. a dust grain at the inner disk rim.
Assuming a very short radiative timescale at the inner rim, no variation in the disk radius and a circular orbit, the temperature of a dust grain ($T_\mathrm{rim}$) ranges from 1400 to 1610\,K.
We can roughly estimate the inner rim scale-height ($H$) due to hydro-static equilibrium in the case of a central binary with:
\begin{equation}
    H = \sqrt{\frac{kT_\mathrm{rim}}{\mu_\mathrm{g}G}\Bigg(\frac{M_1}{d_1^3} + \frac{M_2}{d_2^3} \Bigg)^{-1}}
\end{equation}
where $k$ is the Boltzmann constant, $\mu_\mathrm{g}\simeq2.3\,m_\mathrm{p}$ is the mean molecular weight, $m_\mathrm{p}$ is the proton mass, $G$ is the gravitational constant, $d_1$ the distance to the primary and $d_2$ the distance to the secondary.
We can compute the rim scale-height for the two extreme orbital phases w.r.t. a dust particle at the inner rim: where the primary is the closest to the dust particle and the point where it is the furthest.
This will translate into a variation of the scale-height of the inner rim of \modif{2.4}\% (from 1.1\modif{5}\,au to 1.\modif{18}\,au).
This can explain only part of the brightness luminosity variations we see in the actual interferometric data.
In the polar plot we see a variation of 91\% of the disk brightness.
Moreover, these disk scale-height variations due to the hydro-static equilibrium are expected to be smooth throughout the disk.
The image shows a steep decrease in brightness between the maximum at 195$^\circ$ and the minimum at 125$^\circ$.
We conclude that this kind of feature cannot be reproduced by a change in the local hydro-static equilibrium alone.

\subsection{Origin of the extended flux}

Our radiative transfer model alone could not account for the whole over-resolved flux (15.5\% at 1.65$\mu$m; Paper~I) as we needed to artificially add 8.1\% of over-resolved flux.
It means that the model is reproducing approximately half of the over-resolved flux by scattering the stellar flux on the disk surface.
The problem of an unaccounted over-resolved flux contribution was also found when modelling 89~Her \citep{2014aahillen} but it was detected at visible wavelengths.
The origin of this high level of over-resolved flux in our target is not clear.

One possibility is that it comes from a more complex disk structure.
For example a spiral or a gaped disk can produce more extended emission in the near-infrared \citep[e.g.][]{Fukagawa2010,Tatulli2011}.
In gaped disks around YSOs, the polycyclic aromatic hydrocarbons (PAHs) in the gap are directly exposed to stellar high-energy photons from the central star.
These PAHs are quantum heated and can radiate in the near-infrared continuum \citep{Klarmann2017} explaining the over-resolved flux seen in near-infrared interferometric data of some of the young targets.
However, IRAS08544-4431 is an oxygen-rich evolved star in which the circumstellar gas and dust originates from the star itself and therefore the disk is not expected to show PAH-features.
Moreover, the central star is not emitting a great amount of high-energy photons, necessary to heat the PAHs to near-infrared emitting temperatures.

\begin{figure}[!t]
    \centering
    \includegraphics[width=10cm]{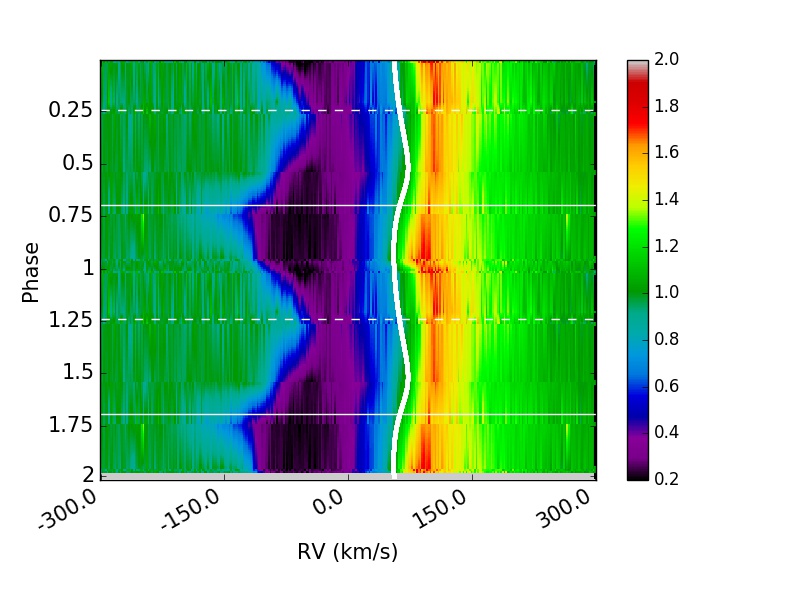}
    \caption{Dynamical spectrum in H$_\alpha$ as function of the orbital phase. The x-axis is the radial velocity and the y-axis is the orbital phase.
    The white solid curve indicate the radial velocity of the primary star.
    The solid and dashed horizontal lines indicate the superior and inferior conjunctions respectively.
    }
    \label{fig:interpol}
\end{figure}

Another possibility is that the extended component is coming from the primary flux scattered by dust in the outflow of the secondary.
This outflow is not modeled by our 2D dust disk model.
The secondary is surrounded by an accretion disk and launches a wide wind \citep{2003aamaas}.
If this wind is loaded with dust, it can emit in the near-infrared continuum and be very extended \citep[e.g.][]{Bans2012}.
One way to have dust in the wind is that it is brought by the accretion disk.
However, it is unlikely that dust particles can survive inside the dust sublimation radius of 8\,au without sublimating.
Another way is that the wind takes the dust from the disk inner regions.
The H$_\alpha$ profiles in the CORALIE spectra always show a P-Cygni profile with an absorption feature which means that the outflow opening angle is at least as large as the disk inclination ($i>20^\circ$; see Fig.\,\ref{fig:interpol}).
The opening angle should therefore be larger than 20$^\circ$.
An opening angle of 70$^\circ$ would reach the disk atmosphere 5\,au above the disk midplane at 15\,au from the mass centre.
At this distance from the midplane the dust density in this upper disk layer is 1000$\times$ smaller than in the midplane ($\rho_\mathrm{dust}\sim10^{-16}$g/cm$^3$). 
Whether or not this amount of dust is able to scatter the primary flux and account for the $\sim$8\% of over-resolved emission has to be answered by further modeling work.

Therefore, the origin of the extended flux is not clear and further observations are needed to characterize it like direct imaging with an instrument with adaptive optics and/or aperture masking observations.

\subsection{Comparison to YSOs}

We have successfully reproduced the radial structure of the disk with a radiative transfer model developed to reproduce disks around YSOs.
For disks around YSOs these models have trouble to reproduce the near-infrared flux and interferometric data simultaneously.
A protoplanetary disk with hydrostatic equilibrium will not produce enough near-infrared emission at the inner rim.
The inner rim emission is radially extended \citep{Lazareff2017} probably due to dust segregation \citep{Tannirkulam2007,Kama2009}, disk accretion \citep{Flock2016} and/or the presence of a thick gaseous disk inside the dust sublimation radius \citep{Kraus2008}.

A rounded inner rim at hydro-static equilibrium is enough to reproduce both the SED and the interferometric observables around IRAS08544-4431.
There is no need to add an emission inside the dust sublimation radius.
This disk rim is also radially sharp compared to disk rims around young stellar objects \citep[e.g.][]{Lazareff2017}.
The photosphere of IRAS08544-4431 is affected by depletion \citep{2003aamaas} suggesting that accretion from the circumbinary disk occured or is occuring.
The gas/dust separation is expected for accretion rates smaller than 10$^{-6}$M$_\odot$/yr \citep{Waters1992}.
\citet{Flock2016} show that disks around YSOs show a radially extended rim for accretion rates down to 10$^{-9}$.
The radial extension of the disk seems more dependent on the luminosity of the central star, the inner rim being sharper at higher stellar luminosities. 
However, \citet{Flock2016} computed models for YSOs and did not extrapolate above a luminosity of 56\,L$_\odot$.
For a luminosity of 14000\,L$_\odot$ the inner rim profile could therefore be more vertical even for moderate accretion rates.

\section{Conclusions}
\label{sec:ccl}

We present in this paper a radiative transfer model of the circumbinary disk around the post-AGB binary IRAS08544-4431.
We successfully reproduced the SED and the radial structure of the disk inner rim by reproducing the squared visibility measurements.
We used a classical self-consistent model designed to reproduce disks around YSOs by defining the disk vertical scale-height by computing the hydrostatic equilibrium.
Whereas the model has moderate success in reproducing disks around YSOs it turns out that it is successful to reproduce circumbinary disks around post-AGBs.

The main results of this paper are:
\begin{itemize}
\item the inner disk rim is ruled by dust sublimation physics and is \modif{well reproduced by our model which is} in hydro-static equilibrium.
\item the inner disk rim is not axi-symmetric. This asymmetry might be explained by the central binary orbit but a detailed modeling of its effects is required to see whether it can truly explain the steep variations in azimuthal brightness at the disk rim.
\item an over-resolved near-infrared component is present and cannot be reproduced by a pure disk model. Its origin remains unclear but is likely linked to the low-mass outflow which is present in this system as well, as evidenced by the H$_\alpha$ profile.
\end{itemize}

Given the number of post-AGB binaries with circumbinary disks, it is clear that complex disk structures are indeed common around post-AGB binaries. 
To answer the questions of 1) the physics ruling the radius of the inner rim (dust sublimation? binary dynamical truncation?), 2) the physics of the inner rim perturbation (what is the influence of the inner binary?), 3) the systematic detection of an emission around the secondary star (are circumcompanion accretion disks common and how are they fed), 4) the origin of the extended emission (disk structure? disk wind?) and 5) the physical relation between the compact dust disk as observed in the near-IR and the large gaseous disk as detected by ALMA,  a systematic study of the near-infrared emission as well as the extended emission around post-AGB binaries is needed. To study the disk-binary interaction, we are planning to monitor this system with time-resolved near-infrared long baseline interferometric observations.

We conclude that in many ways the disks around luminous post-AGB binaries are scaled-up, more irradiated versions of protoplanetary disks around YSOs.

\begin{acknowledgements}
JK and HVW acknowledge support from the Research Council of the KU Leuven under grant number C14/17/082. RM acknowledges support from the Research Council of the KU Leuven under contract GOA/13/012 and the Belgian Science Policy Office under contract BR/143/A2/STARLAB. We used the following internet-based resources: NASA Astrophysics Data System for bibliographic services; Simbad; the VizieR online catalogues operated by CDS.
\end{acknowledgements}

\bibliographystyle{aa} % style aa.bst
\bibliography{biblio} % your references Yourfile.bib

\begin{appendix}

\section{Photometry}
\label{app:sed}

\begin{table*}[!th]
\caption{Photometry of IRAS~08544-4431 \label{tab:SEDapp}}
\begin{center}
\begin{tabular}{rrrrr}
$\lambda_\mathrm{eff}$ & Photometric band & $\lambda$ F$_\mathrm{lambda}$ & error &Ref. \\
\modif{m} & - & \modif{W.m$^{-2}$} & \modif{W.m$^{-2}$}  &- \\
\hline
%\startdata
3.46e-07 & Johnson:U & 8.25e-13 & 0.23e-13 & \modif{(1)} \\
3.46e-07 & Johnson:U & 6.14e-13 & 0.16e-13 & \modif{(1)} \\
4.42e-07 & Johnson:B & 4.07e-12 & 0.11e-12 & \modif{(1)} \\
4.42e-07 & Johnson:B & 3.29e-12 & 0.09e-12 & \modif{(1)} \\
4.42e-07 & Johnson:B & 4.04e-12 & 0.07e-12 & \modif{(2)} \\
4.42e-07 & TYCHO2:BT & 2.70e-12 & 0.41e-12 & \modif{(3)} \\
5.4e-07 & Johnson:V & 8.68e-12 & 0.24e-12 & \modif{(1)} \\
5.4e-07 & Johnson:V & 7.42e-12 & 0.20e-12 & \modif{(1)} \\
5.4e-07 & Johnson:V & 7.53e-12 & 0.13e-12 & \modif{(2)} \\
5.4e-07 & TYCHO2:VT & 6.49e-12 & 0.11e-12 & \modif{(3)} \\
6.47e-07 & Cousins:R & 1.15e-11 & 0.03e-11 & \modif{(1)} \\
6.47e-07 & Cousins:R & 1.01e-11 & 0.03e-11 & \modif{(1)} \\
7.865e-07 & Cousins:I & 1.64e-11 & 0.06e-11 & \modif{(1)} \\
7.865e-07 & Cousins:I & 1.47e-11 & 0.04e-11 & \modif{(1)} \\
1.25e-06 & 2MASS:J & 1.89e-11 & 0.04e-11 & \modif{(4)} \\
 1.2e-06 & SAAO:J & 1.96e-11 & 0.05e-11 & \modif{(1)} \\
1.65e-06 & 2MASS:H & 4.743 & 0.040 & \modif{(4)} \\
 1.6e-06& SAAO:H & 4.67 & 0.03 & \modif{(1)} \\
  2.2e-06& SAAO:K & 3.51 & 0.03 & \modif{(1)} \\
2.15e-06 & 2MASS:Ks & 3.523 & 0.244 & \modif{(4)} \\
3.5e-06& SAAO:L & 1.59 & 0.03 & \modif{(1)}\\
4.3e-06 & MSX:B1 & 76.62Jy & 6.74 & \modif{(5)} \\
4.4e-06 & MSX:B2 & 51.92Jy & 4.78 & \modif{(5)} \\
8.3e-06 & MSX:A & 139.0Jy & 5.7 & \modif{(5)} \\
9e-06 & AKARI:S9W & 156.6Jy & 4.2 & \modif{(6)} \\
1.2e-05 & MSX:C & 150.7Jy & 7.5 & \modif{(5)} \\
1.2e-05 & IRAS:F12 & 186.6Jy & 9.0 & \modif{(7)} \\
1.2e-05 & WISE:W3 & -1.822 & 0.276 & \modif{(8)} \\
1.5e-05 & MSX:D & 143.6Jy & 8.8 & \modif{(5)} \\
1.8e-5 & AKARI:L18W & 155.9Jy & 1.4 & \modif{(6)} \\
2.1e-05 & MSX:E & 134.2Jy & 8.1 & \modif{(5)} \\
2.2e-05 & WISE:W4 & -3.116 & 0.001 & \modif{(8)} \\
2.5e-05 & IRAS:F25 & 152.7Jy & 7.9 & \modif{(7)} \\
6e-05 & IRAS:F60 & 53.78Jy & 2.80 & \modif{(7)} \\
6.5e-05 & AKARI:N60 & 45.66Jy & 1.33 & \modif{(9)} \\
9e-05 & AKARI:WIDES & 27.06Jy & 1.96 & \modif{(9)} \\
0.00014 & AKARI:WIDEL & 7.342Jy & 1.04 & \modif{(9)} 
\end{tabular}
\tablebib{\modif{(1) \citet{deruyter06}, (2) \citet{Kharchenko2009}, (3) \citet{Hog2000}, (4) \citet{2003ycatcutri}, (5) \citet{Egan2003}, (6) \citet{Ishihara2010}, (7) \citet{Moshir2008}, (8) \citet{2012ycatcutri}, (9) \citet{Yamamura2010}.
}}
\end{center}
%\enddata
%\tablecomments{Note that {\tt \string \colnumbers} does not work with the 
%vertical line alignment token. If you want vertical lines in the headers you
%can not use this command at this time.}
\end{table*}

\begin{figure*}[!th]
    \centering
    \includegraphics[width=7cm]{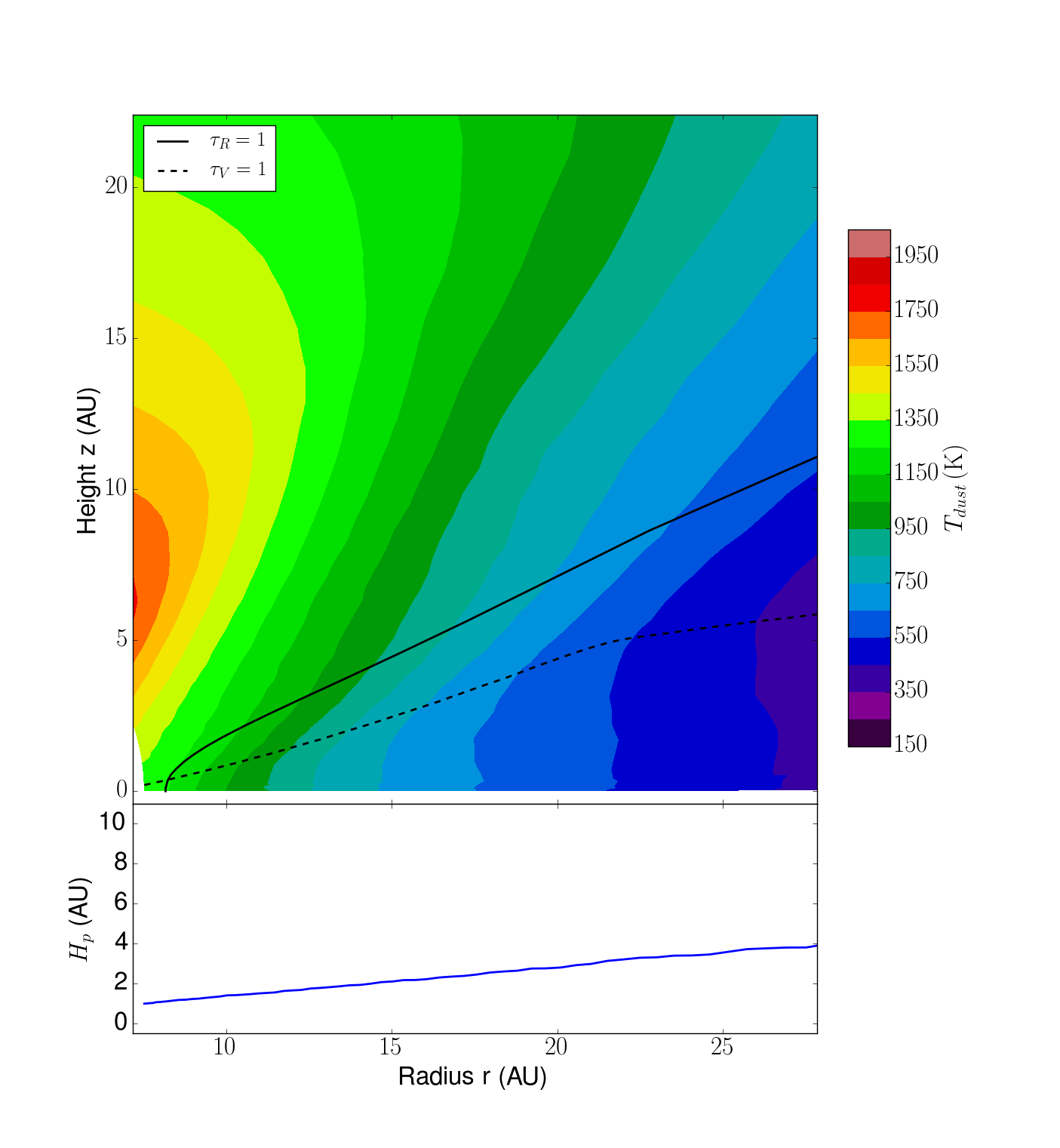}
    \includegraphics[width=6cm]{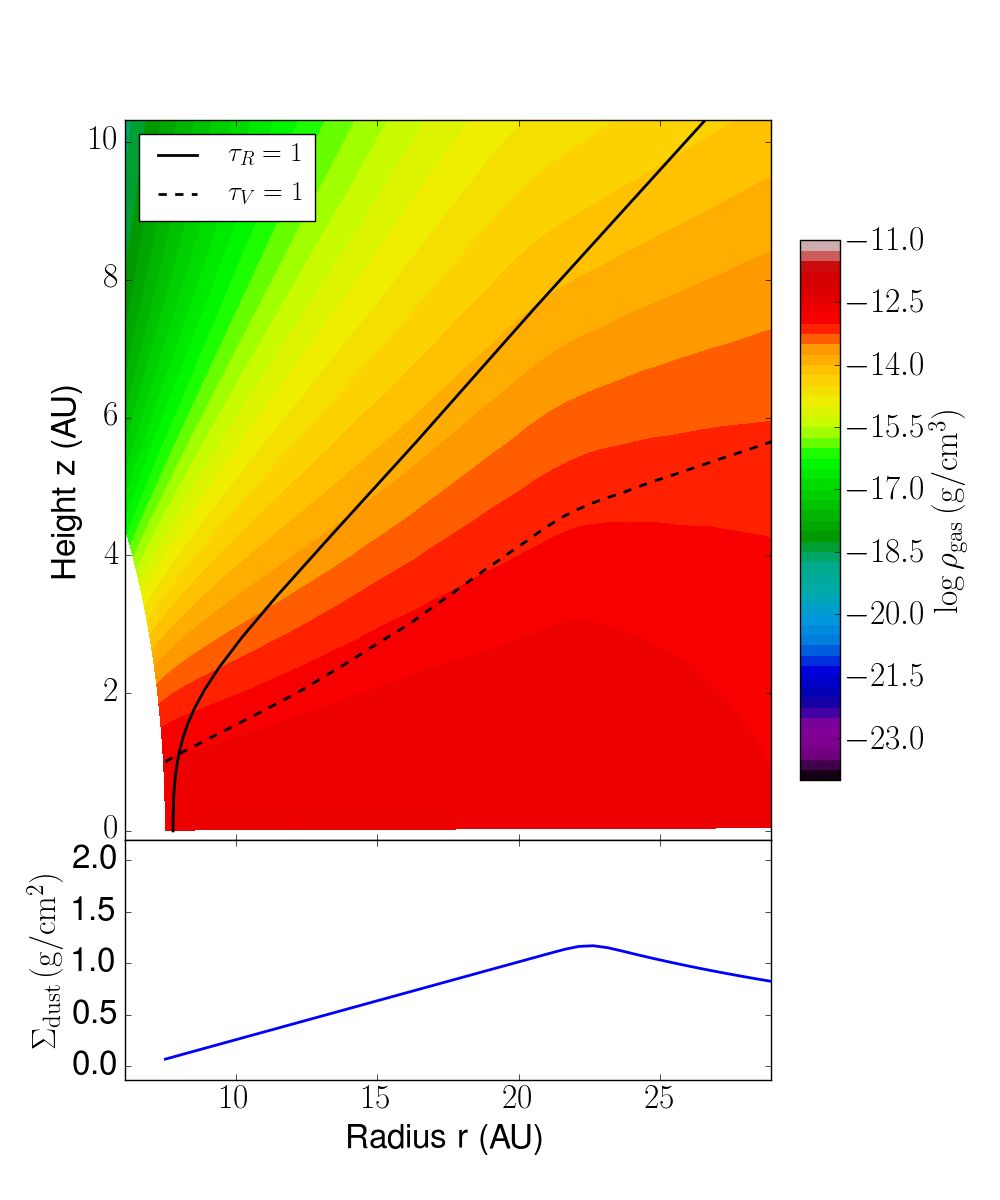}
    \caption{Best fit radiative transfer model. Left: the temperature in the model. Right: the gas density in the model.
    The lines represent the radial (solid) and vertical (dashed) $\tau$=1 surface.}
    \label{fig:Temp}
\end{figure*}

\end{appendix}

\end{document}